\documentclass[a4paper,twosided,notoc]{tufte-book}
\usepackage{siunitx}
\hypersetup{colorlinks}

\title{tikz-network\\manual}
\author[J\"urgen Hackl]{J\"urgen Hackl}
\publisher{Version 1.0}

\usepackage{framed}
\definecolor{shadecolor}{HTML}{abd7e6}
\definecolor{hlcolor}{HTML}{057498}
\usepackage{tikz}
\usetikzlibrary{backgrounds}
\tikzset{background rectangle/.style={fill=yellow!30}}

\usepackage{lipsum}

\usepackage{booktabs}
\usepackage{tabularx}

\usepackage{graphicx}
\setkeys{Gin}{width=\linewidth,totalheight=\textheight,keepaspectratio}
\graphicspath{{graphics/}}

\usepackage{fancyvrb}
\fvset{fontsize=\normalsize}

\usepackage{xspace}

\newcommand{\monthyear}{%
  \ifcase\month\or January\or February\or March\or April\or May\or June\or
  July\or August\or September\or October\or November\or
  December\fi\space\number\year
}

\usepackage{units}

\newcommand{\hlred}[1]{\textcolor{hlcolor}{#1}}
\newcommand{\hangleft}[1]{\makebox[0pt][r]{#1}}
\newcommand{\TODO}{\textcolor{red}{\bf TODO!}\xspace}
\providecommand{\XeLaTeX}{X\lower.5ex\hbox{\kern-0.15em\reflectbox{E}}\kern-0.1em\LaTeX}

\newcommand{\tuftebs}{\symbol{'134}}
\newcommand{\doccmddef}[2][]{%
  \hlred{\texttt{\tuftebs#2}}\label{cmd:#2}%
  \ifthenelse{\isempty{#1}}%
    {
      \index{#2 command@\protect\hangleft{\texttt{\tuftebs}}\texttt{#2}}
    }%
    {
      \index{#2 command@\protect\hangleft{\texttt{\tuftebs}}\texttt{#2} (\texttt{#1} package)}
      \index{#1 package@\texttt{#1} package}\index{packages!#1@\texttt{#1}}
    }%
}
\newcommand{\doccmd}[2][]{%
  \texttt{\tuftebs#2}%
  \ifthenelse{\isempty{#1}}%
    {
      \index{#2 command@\protect\hangleft{\texttt{\tuftebs}}\texttt{#2}}
    }%
    {
      \index{#2 command@\protect\hangleft{\texttt{\tuftebs}}\texttt{#2} (\texttt{#1} package)}
      \index{#1 package@\texttt{#1} package}\index{packages!#1@\texttt{#1}}
    }%
}
\newcommand{\docopt}[1]{\ensuremath{\langle}\textrm{\textit{#1}}\ensuremath{\rangle}}
\newcommand{\docarg}[1]{\textrm{\textit{#1}}}
\newenvironment{docspecdef}{\begin{quotation}\ttfamily\parskip0pt\parindent0pt\ignorespaces}{\end{quotation}}
\newcommand{\docenv}[1]{\texttt{#1}\index{#1 environment@\texttt{#1} environment}\index{environments!#1@\texttt{#1}}}
\newcommand{\docenvdef}[1]{\hlred{\texttt{#1}}\label{env:#1}\index{#1 environment@\texttt{#1} environment}\index{environments!#1@\texttt{#1}}}
\newcommand{\doccls}[1]{\texttt{#1}}
\newcommand{\docclsopt}[1]{\texttt{#1}\index{#1 class option@\texttt{#1} class option}\index{class options!#1@\texttt{#1}}}
\newcommand{\docmsg}[2]{\bigskip\begin{fullwidth}\noindent\ttfamily#1\end{fullwidth}\medskip\par\noindent#2}

\newenvironment{docspec}{\begin{shaded}}{\vspace{-5mm}\end{shaded}}

\usepackage{etoolbox}

\makeatletter
\patchcmd{\ttlh@hang}{\parindent\z@}{\parindent\z@\leavevmode}{}{}
\patchcmd{\ttlh@hang}{\noindent}{}{}{}
\makeatother

\titleformat{\chapter}%
  {\huge\rmfamily\itshape}
  {\llap{\parbox{1.5cm}{\hfill\itshape\huge\thechapter}\hspace{2mm}}}
  {0em}
  {}
  []

\titleformat{\section}%
  {\normalfont\Large\itshape}
  {\llap{\parbox{1.0cm}{\hfill\thesection}\hspace{2mm}}}
  {0em}
  {}
  []

\titleformat{\subsection}%
  {\normalfont\large\itshape}
  {\llap{\parbox{1.5cm}{\hfill\thesubsection}\hspace{2mm}}}
  {0em}
  {}
  []

\newcommand{\pkg}{\doccls{tikz-network}\xspace}
\newcommand{\tikzsym}{Ti\emph{k}Z }

\usepackage{listings}
\usepackage{tikz-network}

\usepackage{makeidx}
\makeindex

\begin{document}


\thispagestyle{empty}

  \begin{fullwidth}%
\sffamily%
  \fontsize{18}{20}\selectfont\par\noindent\textcolor{darkgray}{\allcaps{\thanklessauthor}}%
  \vspace{11.5pc}%
  \fontsize{36}{40}\selectfont\par\noindent\textcolor{darkgray}{\allcaps{\thanklesstitle}}%
\vspace{3cm}
\begin{center}
\includegraphics[width=14cm]{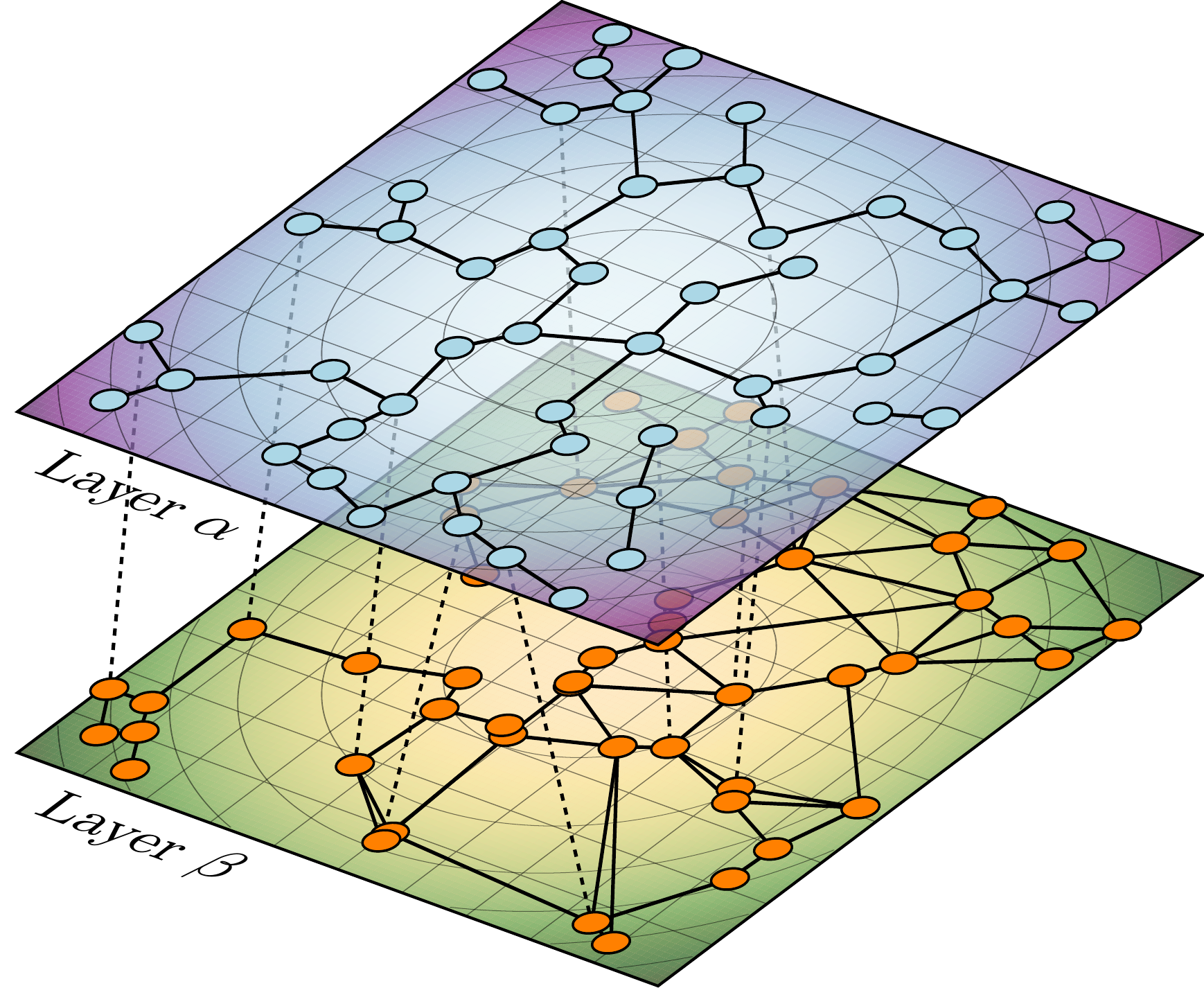}%
\end{center}
\vfill%
  \fontsize{14}{16}\selectfont\par\noindent\allcaps{\thanklesspublisher}%
  \end{fullwidth}%

\newpage

\begin{fullwidth}

~\vfill
\thispagestyle{empty}
\setlength{\parindent}{0pt}
\setlength{\parskip}{\baselineskip}

\par\smallcaps{\pkg}
\par\smallcaps{\thanklesspublisher}
\par\smallcaps{\url{https://github.com/hackl/tikz-network}}

\par\textit{Released, \monthyear}

Copyright \copyright\ \the\year\ \thanklessauthor

This program is free software: you can redistribute it and/or modify
it under the terms of the GNU General Public License as published by
the Free Software Foundation, either version 3 of the License, or
(at your option) any later version.

This program is distributed in the hope that it will be useful,
but WITHOUT ANY WARRANTY; without even the implied warranty of
MERCHANTABILITY or FITNESS FOR A PARTICULAR PURPOSE.  See the
GNU General Public License for more details.

You should have received a copy of the GNU General Public License
along with this program.  If not, see \url{https://www.gnu.org/licenses/}.

\end{fullwidth}

\begin{fullwidth}
\tableofcontents
\end{fullwidth}

\cleardoublepage
\chapter{Introduction}

In recent years, complex network theory becomes more and more popular within the scientific community. Besides a solid mathematical base on which these theories are built on, a visual representation of the networks allow communicating complex relationships to a broad audience.

Nowadays, a variety of great visualization tools are available, which helps to structure, filter, manipulate and of course to visualize the networks. However, they come with some limitations, including the need for specific software tools, difficulties to embed the outputs properly in a \LaTeX~file (e.g. font type, font size, additional equations and math symbols needed, \dots) and challenges in the post-processing of the graphs, without rerunning the software tools again.

In order to overcome this issues, the package \pkg was created. Some of the features are:

\begin{itemize}
\item \LaTeX~ is a standard for scientific publications and widely used
\item beside \LaTeX~ no other software is needed
\item no programming skills are needed
\item simple to use but allows 100\% control over the output
\item easy for post-processing (e.g. adding drawings, texts, equations,\dots)
\item same fonts, font sizes, mathematical symbols, \dots as in the document
\item no quality loss of the output due to the pdf format
\item networks are easy to adapt or modify for lectures or small examples
\item able to visualize larger networks
\item three-dimensional visualization of (multilayer) networks
\item compatible with other visualization tools
\end{itemize}
\newpage

\section{How to read this manual?}

The aim of this manual is to describe the use of the \pkg library for visualizing networks. To ensure an easy use of the elements and to keep the clarity, this manual is structured as follows:
\begin{itemize}
\item In Chapter \ref{chap:simple_networks} the elements to create simple networks (by hand) in a plane are explained. Thereby, the use of the commands \doccmd{Vertex} and \doccmd{Edge} are shown.
\item How to create complex networks from external files\footnote{e.g. \doccls{igraph} or \doccls{networkx}} are explained in Chapter \ref{chap:complex_networks}. The main commands, therefore are \doccmd{Vertices} and \doccmd{Edges} which are using the same options as in the simple case.
\item In Chapter \ref{chap:multilayer_networks}, the visualization of multilayer networks is explained. Additional visualization tools such as \doccmd{Plane} and \docenv{Layer} are introduced.
\item The default settings used and how they can be modified is explained in Chapter \ref{chap:default_settings}.
\item Information about troubleshooting and support is given in Chapter \ref{chap:troubleshooting}
\item Since this is the alpha version (0.1) of the package, features which will be probably added and commands which have to be fixed are listed in Appendix \ref{chap:todo}.
\end{itemize}

\subsection{A few explanations}

The images in this manual are created with the \pkg library or \tikzsym. The code used for this is
specified for each image.
\begin{marginfigure}[30mm]
\centering
  \begin{tikzpicture}
    \filldraw (-.2,.2) circle (2pt) (.2,.2) circle (2pt);
    \draw (0,0) circle (5mm) (-.3,-.1) .. controls (0,-.3) .. (.3,-.1);
  \end{tikzpicture}
\end{marginfigure}

\begin{docspec}
\begin{lstlisting}
  \begin{tikzpicture}
    \filldraw (-.2,.2) circle (2pt) (.2,.2) circle (2pt);
    \draw (0,0) circle (5mm) (-.3,-.1) .. controls (0,-.3) .. (.3,-.1);
  \end{tikzpicture}
\end{lstlisting}
\end{docspec}

Special additions which are needed for a better understanding are shown in orange but are not in the sample code available.

\begin{marginfigure}[23mm]
\centering
  \begin{tikzpicture}
    \filldraw [orange] (0,0) circle (2pt)
    (1,1) circle (2pt)
    (2,1) circle (2pt)
    (2,0) circle (2pt);
    \draw (0,0) .. controls (1,1) and (2,1) .. (2,0);
  \end{tikzpicture}
\end{marginfigure}

\begin{docspec}
\begin{lstlisting}
  \begin{tikzpicture}
    \draw (0,0) .. controls (1,1) and (2,1) .. (2,0);
  \end{tikzpicture}
\end{lstlisting}
\end{docspec}

\subsection{Inputs}
\label{sec:inputs}

The commands in the \pkg library (e.g. \doccmd{Vertex}, \doccmd{Edge}) always start with capital letters and DO NOT need a semicolon <<;>> at the end. Boolean arguments start also with capital letters (e.g. \docopt{NoLabel}). Arguments which need an user input, use are written in small letters (e.g. \docopt{color}).

Basically, one can distinguish between the mandatory argument $\{~\}$ and the optional argument $[~]$. The first values must be entered compulsory. By contrast, nothing has to be entered for the optional input. Additional features (e.g. \docopt{size})) can be activated when entering optional parameters.

When entering size values the base unit is always predefined in $[cm]$\footnote{The default unit can be changed with \doccmd{SetDefaultUnit}; see Section~\ref{sec:gerneral_settings}}, except for line widths which are dedined in $[pt]$. Percentage values $\%$ are always specified as decimal values; for example, $100\% = 1.0 $ and $ 10 \% $ corresponds to $ 0.1 $.

\subsection{Additional help}

Is the manual not enough, occur some ambiguities or some \tikzsym commands are unclear, please have a look in the ``\tikzsym and PGF Manual'' from Till Tantau\footnote{\url{http://mirror.switch.ch/ftp/mirror/tex/graphics/pgf/base/doc/pgfmanual.pdf}}.

Should you have any further questions, please do not hesitate to contact me.

\section{Installation}
\label{sec:Installation}

Actually, we can hardly speak of an installation since only the necessary package \pkg must be loaded in the preamble of your document.

The current release of the package is available via CTAN\footnote{\url{https://ctan.org/tex-archive/graphics/pgf/contrib/tikz-network}}. A release candidate for the next version of \pkg is available on github\footnote{\url{https://github.com/hackl/}}

Is the package installed or the style file is stored in the folder of the main file, so the library can be imported, as the following example shows:

\begin{docspec}
\begin{lstlisting}
  % ------------
  % header
  \documentclass{scrreprt}

item  % ------------
  % packages
  \usepackage{tikz-network}
\end{lstlisting}
\end{docspec}

\section{Additional necessary packages}

To use all commands and options of \tikzsym, possibly some packages need to be reloaded. These missing files (or their names) appear in the error log when you convert the file. However, for the package described in this manual, it is sufficient to use the library and the \tikzsym standard commands.

\chapter{Simple Networks}
\label{chap:simple_networks}

\section{Vertex}
\label{sec:vertex}
On essential command is \doccmddef{Vertex}, which allow placing vertices in the document and modify their appearance.
\begin{docspecdef}
  \doccmd{Vertex}[\docopt{local options}]\{\docarg{Name}\}
\end{docspecdef}
In order to be able to place a vertex, a non-empty \docarg{Name} argument is required. This argument defines the vertex's reference name, which must be unique. Mathematical symbols are not allowed as name as well as no blank spaces. The \docarg{Name} should not be confused with the \docopt{label}, that is used for display; for example one may want to display $A_1$ while the name will be \docarg{A1}.

For a \doccmd{Vertex} the following options are available:

\begin{table}[h]\index{Vertex!options}
  \footnotesize%
  \begin{center}
    \begin{tabular}{lccl}
      \toprule
      Option & Default & Type &Definition \\
      \midrule
      x          & 0     & measure& x-coordinate \\
      y          & 0     & measure& y-coordinate \\
      size       & \{\}  & measure& diameter of the circle\\
      color      & \{\}  & color & fill color of vertex\\
      opacity    & \{\}  & number & opacity of the fill color \\
      shape      & \{\}  & string & shape of the vertex\\
      label      & \{\}  & string & label \\
      fontsize   & \{\}  & string & font size of the label\\
      fontcolor  & \{\}  & color & font color of the label\\
      fontscale  & \{\}  & number & scale of the label\\
      position   & center& value$^a$ & label position\\
      distance   & 0     & measure & label distance from the center \\
      style      & \{\}  & string & additional \tikzsym styles \\
      layer      & \{\}  & number & assigned layer of the vertex \\
      \midrule
      NoLabel    & false & Boolean & delete the label \\
      IdAsLabel  & false & Boolean & uses the \docarg{Name} as label \\
      Math       & false & Boolean & displays the label in math mode \\
      RGB        & false & Boolean & allow RGB colors \\
      Pseudo     & false & Boolean & create a pseudo vertex\\
      \bottomrule
    \end{tabular}
    \scriptsize
    \\$^a$ either measure or string
  \end{center}
  \caption{Local options for the \doccmd{Vertex} command.}
  \label{tab:vertex_options}
\end{table}

The order how the options are entered does not matter. Changes to the default Vertex layout can be made with \doccmd{SetVertexStyle}\footnote{see Section \ref{sec:vertex_style}}

\begin{docspecdef}
  \doccmd{Vertex}[\docopt{x}=\docarg{measure},\docopt{y}=\docarg{measure}]\{\docarg{Name}\}
\end{docspecdef}

The location of the vertices are determined by Cartesian coordinates in \docopt{x} and \docopt{y}. The coordinates are optional. If no coordinates are determined the vertex will be placed at the origin ($0,0$). The entered \docarg{measures} are in default units (\si{cm}). Changing the unites (locally) can be done by adding the unit to the \docarg{measure}\footnote{e.g. x=\SI{1}{in}}. Changes to the default setting can be made with \doccmd{SetDefaultUnit}\footnote{see Section \ref{sec:gerneral_settings}}.

\begin{marginfigure}[6mm]
\centering
  \begin{tikzpicture}
    \Vertex{A}
    \Vertex[x=1,y=1]{B}
    \Vertex[x=2]{C}
    \node at (0,0)[font=\footnotesize,orange]{\textbf{A}};
    \node at (1,1)[font=\footnotesize,orange]{\textbf{B}};
    \node at (2,0)[font=\footnotesize,orange]{\textbf{C}};
  \end{tikzpicture}
\end{marginfigure}

\begin{docspec}
\begin{lstlisting}
  \begin{tikzpicture}
    \Vertex{A}
    \Vertex[x=1,y=1]{B}
    \Vertex[x=2]{C}
  \end{tikzpicture}
\end{lstlisting}
\end{docspec}

\begin{docspecdef}
  \doccmd{Vertex}[\docopt{size}=\docarg{measure}]\{\docarg{Name}\}
\end{docspecdef}

The diameter of the vertex can be changed with the option \docopt{size}. Per default a vertex has \SI{0.6}{cm} in diameter. Also, here the default units are \si{cm} and have not to be added to the \docarg{measure}.

\begin{marginfigure}[34mm]
\centering
  \begin{tikzpicture}
    \Vertex[size=.3]{A}
    \Vertex[x=1,size=.7]{B}
    \Vertex[x=2.3,size=1]{C}
  \end{tikzpicture}
\end{marginfigure}

\begin{docspec}
\begin{lstlisting}
  \begin{tikzpicture}
    \Vertex[size=.3]{A}
    \Vertex[x=1,size=.7]{B}
    \Vertex[x=2.3,size=1]{C}
  \end{tikzpicture}
\end{lstlisting}
\end{docspec}

\begin{docspecdef}
  \doccmd{Vertex}[\docopt{color}=\docarg{color}]\{\docarg{Name}\}
\end{docspecdef}

To change the fill color of each vertex individually, the option \docopt{color} has to be used. Without the option \docopt{RGB} set, the default \tikzsym and \LaTeX~ colors can be applied.

\begin{marginfigure}[34mm]
\centering
  \begin{tikzpicture}
    \Vertex[color = blue]{A}
    \Vertex[x=1,color=red]{B}
    \Vertex[x=2,color=green!70!blue]{C}
  \end{tikzpicture}
\end{marginfigure}

\begin{docspec}
\begin{lstlisting}
  \begin{tikzpicture}
    \Vertex[color = blue]{A}
    \Vertex[x=1,color=red]{B}
    \Vertex[x=2,color=green!70!blue]{C}
  \end{tikzpicture}
\end{lstlisting}
\end{docspec}

\begin{docspecdef}
  \doccmd{Vertex}[\docopt{opacity}=\docarg{number}]\{\docarg{Name}\}
\end{docspecdef}

With the option \docopt{opacity} the opacity of the vertex fill color can be modified. The range of the \docarg{number} lies between $0$ and $1$. Where $0$ represents a fully transparent fill and $1$ a solid fill.

\begin{marginfigure}[34mm]
\centering
  \begin{tikzpicture}
    \draw[thick,orange,dashed] (-.5,0) -- (2.5,0);
    \Vertex[opacity = 1]{A}
    \Vertex[x=1,opacity =.7]{B}
    \Vertex[x=2,opacity =.2]{C}
  \end{tikzpicture}
\end{marginfigure}

\begin{docspec}
\begin{lstlisting}
  \begin{tikzpicture}
    \Vertex[opacity = 1]{A}
    \Vertex[x=1,opacity =.7]{B}
    \Vertex[x=2,opacity =.2]{C}
  \end{tikzpicture}
\end{lstlisting}
\end{docspec}

\begin{docspecdef}
  \doccmd{Vertex}[\docopt{shape}=\docarg{string}]\{\docarg{Name}\}
\end{docspecdef}

With the option \docopt{shape} the shape of the vertex can be modified. Thereby the shapes implemented in \tikzsym can be used, including: \docarg{circle}, \docarg{rectangle}, \docarg{diamond}, \docarg{trapezium}, \docarg{semicircle}, \docarg{isosceles triangle}, \dots.

\begin{marginfigure}[34mm]
\centering
  \begin{tikzpicture}
    \Vertex[shape = rectangle]{A}
    \Vertex[x=1,shape = diamond]{B}
    \Vertex[x=2,shape = isosceles triangle]{C}
  \end{tikzpicture}
\end{marginfigure}

\begin{docspec}
\begin{lstlisting}
  \begin{tikzpicture}
    \Vertex[shape = rectangle]{A}
    \Vertex[x=1,shape = diamond]{B}
    \Vertex[x=2,shape = isosceles triangle]{C}
  \end{tikzpicture}
\end{lstlisting}
\end{docspec}

\begin{docspecdef}
  \doccmd{Vertex}[\docopt{label}=\docarg{string}]\{\docarg{Name}\}
\end{docspecdef}

In \pkg there are several ways to define the labels of the vertices and edges. The common way is via the option \docopt{label}. Here, any \docarg{string} argument can be used, including blank spaces. The environment \$ \$ can be used to display mathematical expressions.

\begin{marginfigure}[34mm]
\centering
  \begin{tikzpicture}
    \Vertex[label=foo]{A}
    \Vertex[x=1,label=bar]{B}
    \Vertex[x=2,label=$u_1$]{C}
  \end{tikzpicture}
\end{marginfigure}

\begin{docspec}
\begin{lstlisting}
  \begin{tikzpicture}
    \Vertex[label=foo]{A}
    \Vertex[x=1,label=bar]{B}
    \Vertex[x=2,label=$u_1$]{C}
  \end{tikzpicture}
\end{lstlisting}
\end{docspec}

\begin{docspecdef}
  \doccmd{Vertex}[\docopt{label}=\docarg{string},\docopt{fontsize}=\docarg{string}]\{\docarg{Name}\}
\end{docspecdef}

The font size of the \docopt{label} can be modified with the option \docopt{fontsize}. Here common \LaTeX~ font size commands\footnote{e.g. \doccmd{tiny}, \doccmd{scriptsize}, \doccmd{footnotesize}, \doccmd{small}, \dots.} can be used to change the size of the label.

\begin{marginfigure}[10mm]
\centering
  \begin{tikzpicture}
    \Vertex[label=foo,fontsize=\normalsize]{A}
    \Vertex[x=1,label=bar,fontsize=\tiny]{B}
    \Vertex[x=2,label=$u_1$,fontsize=\large]{C}
  \end{tikzpicture}
\end{marginfigure}

\begin{docspec}
\begin{lstlisting}
  \begin{tikzpicture}
    \Vertex[label=foo,fontsize=\normalsize]{A}
    \Vertex[x=1,label=bar,fontsize=\tiny]{B}
    \Vertex[x=2,label=$u_1$,fontsize=\large]{C}
  \end{tikzpicture}
\end{lstlisting}
\end{docspec}

\begin{docspecdef}
  \doccmd{Vertex}[\docopt{label}=\docarg{string},\docopt{fontcolor}=\docarg{color}]\{\docarg{Name}\}
\end{docspecdef}

The color of the \docopt{label} can be changed with the option \docopt{fontcolor}. Currently, only the default \tikzsym and \LaTeX~ colors are supported\footnote{\TODO Add RGB option!}.

\begin{marginfigure}[8mm]
\centering
  \begin{tikzpicture}
    \Vertex[label=foo,fontcolor=blue]{A}
    \Vertex[x=1,label=bar,fontcolor=magenta]{B}
    \Vertex[x=2,label=$u_1$,fontcolor=red]{C}
  \end{tikzpicture}
\end{marginfigure}

\begin{docspec}
\begin{lstlisting}
  \begin{tikzpicture}
    \Vertex[label=foo,fontcolor=blue]{A}
    \Vertex[x=1,label=bar,fontcolor=magenta]{B}
    \Vertex[x=2,label=$u_1$,fontcolor=red]{C}
  \end{tikzpicture}
\end{lstlisting}
\end{docspec}

\begin{docspecdef}
  \doccmd{Vertex}[\docopt{label}=\docarg{string},\docopt{fontscale}=\docarg{number}]\{\docarg{Name}\}
\end{docspecdef}

Contrary to the option \docopt{fontsize}, the option \docopt{fontscale} does not change the font size itself but scales the curent font size up or down. The \docarg{number} defines the scale, where numbers between $0$ and $1$ down scale the font and numbers greater then $1$ up scale the label. For example $0.5$ reduces the size of the font to $50\%$ of its originial size, while $1.2$ scales the font to $120\%$.

\begin{marginfigure}[34mm]
\centering
  \begin{tikzpicture}
    \Vertex[label=foo,fontscale=0.5]{A}
    \Vertex[x=1,label=bar,fontscale=1]{B}
    \Vertex[x=2,label=$u_1$,fontscale=2]{C}
  \end{tikzpicture}
\end{marginfigure}

\begin{docspec}
\begin{lstlisting}
  \begin{tikzpicture}
    \Vertex[label=foo,fontscale=0.5]{A}
    \Vertex[x=1,label=bar,fontscale=1]{B}
    \Vertex[x=2,label=$u_1$,fontscale=2]{C}
  \end{tikzpicture}
\end{lstlisting}
\end{docspec}

\begin{docspecdef}
  \doccmd{Vertex}[\docopt{label}=\docarg{string},\docopt{position}=\docarg{value},\docopt{distance}=\docarg{number}]\{\docarg{Name}\}
\end{docspecdef}

Per default the \docopt{position} of the \docopt{label} is in the \docarg{center} of the vertex. Classical \tikzsym commands\footnote{e.g. \docarg{above}, \docarg{below}, \docarg{left}, \docarg{right}, \docarg{above left}, \docarg{above right},\dots} can be used to change the \docopt{position} of the \docopt{label}. Instead, using such command, the position can be determined via an angle, by entering a \docarg{number} between $-360$ and $360$. The origin ($0^\circ$) is the $y$ axis. A positive \docarg{number} change the \docopt{position} counter clockwise, while a negative \docarg{number} make changes clockwise.

With the option, \docopt{distance} the distance between the vertex and the label can be changed.

\begin{marginfigure}[31mm]
\centering
  \begin{tikzpicture}
    \Vertex[label=A,position=below]{A}
    \Vertex[x=1,label=B,position=below,distance=2mm]{B}
    \Vertex[x=2,label=C,position=30,distance=1mm]{C}
    \draw[orange,dashed](2,0) --++ (7mm,0mm)(2,0)--++(30:7mm);
    \draw[orange,->] (2.5,0) arc (0:30:5mm);
    \node[orange] (x) at (2.6,-.2) {$30^\circ$};
  \end{tikzpicture}
\end{marginfigure}

\begin{docspec}
\begin{lstlisting}
  \begin{tikzpicture}
    \Vertex[label=A,position=below]{A}
    \Vertex[x=1,label=B,position=below,distance=2mm]{B}
    \Vertex[x=2,label=C,position=30,distance=1mm]{C}
  \end{tikzpicture}
\end{lstlisting}
\end{docspec}

\begin{docspecdef}
  \doccmd{Vertex}[\docopt{style}=\{\docarg{string}\}]\{\docarg{Name}\}
\end{docspecdef}

Any other \tikzsym style option or command can be entered via the option \docopt{style}. Most of these commands can be found in the ``\tikzsym and PGF Manual''. Contain the commands additional options (e.g.\docopt{shading}=\docarg{ball}), then the argument for the \docopt{style} has to be between $\{~\}$ brackets.

\begin{marginfigure}[32mm]
\centering
  \begin{tikzpicture}
    \Vertex[style={color=green}]{A}
    \Vertex[x=1,style=dashed]{B}
    \Vertex[x=2,style={shading=ball}]{C}
  \end{tikzpicture}
\end{marginfigure}

\begin{docspec}
\begin{lstlisting}
  \begin{tikzpicture}
    \Vertex[style={color=green}]{A}
    \Vertex[x=1,style=dashed]{B}
    \Vertex[x=2,style={shading=ball}]{C}
  \end{tikzpicture}
\end{lstlisting}
\end{docspec}

\begin{docspecdef}
  \doccmd{Vertex}[\docopt{IdAsLabel}]\{\docarg{Name}\}
\end{docspecdef}
\begin{docspecdef}
  \doccmd{Vertex}[\docopt{NoLabel},\docopt{label}=\docarg{string}]\{\docarg{Name}\}
\end{docspecdef}

\docopt{IdAsLabel} is a Boolean option which assigns the \docarg{Name} of the vertex as label. On the contrary, \docopt{NoLabel} suppress all labels.

\begin{marginfigure}[32mm]
\centering
  \begin{tikzpicture}
    \Vertex[IdAsLabel]{A}
    \Vertex[x=1,label=B,NoLabel]{B}
    \Vertex[x=2,IdAsLabel,NoLabel]{C}
  \end{tikzpicture}
\end{marginfigure}

\begin{docspec}
\begin{lstlisting}
  \begin{tikzpicture}
    \Vertex[IdAsLabel]{A}
    \Vertex[x=1,label=B,NoLabel]{B}
    \Vertex[x=2,IdAsLabel,NoLabel]{C}
  \end{tikzpicture}
\end{lstlisting}
\end{docspec}

\begin{docspecdef}
  \doccmd{Vertex}[\docopt{Math},\docopt{label}=\docarg{string}]\{\docarg{Name}\}
\end{docspecdef}

The option \docopt{Math} allows transforming labels into mathematical expressions without using the \$~\$ environment. In combination with \docopt{IdAsLabel} allows this option also mathematical expressions by the definition of the vertex \docarg{Name}.

\begin{marginfigure}[32mm]
\centering
  \begin{tikzpicture}
    \Vertex[IdAsLabel]{A1}
    \Vertex[x=1,label=B_1,Math]{B}
    \Vertex[x=2,Math,IdAsLabel]{C_1}
  \end{tikzpicture}
\end{marginfigure}

\begin{docspec}
\begin{lstlisting}
  \begin{tikzpicture}
    \Vertex[IdAsLabel]{A1}
    \Vertex[x=1,label=B_1,Math]{B}
    \Vertex[x=2,Math,IdAsLabel]{C_1}
  \end{tikzpicture}
\end{lstlisting}
\end{docspec}

\begin{docspecdef}
  \doccmd{Vertex}[\docopt{RGB},\docopt{color}=\docarg{RGB values}]\{\docarg{Name}\}
\end{docspecdef}

In order to display RGB colors for the vertex fill color, the option \docopt{RGB} has to be entered. In combination with this option, the \docopt{color} hast to be a list with the \docarg{RGB
values}, separated by <<,>> and within $\{~\}$.\footnote{e.g. the RGB code for white: $\{255,255,255\}$}

\begin{marginfigure}[4mm]
\centering
  \begin{tikzpicture}
    \Vertex[RGB,color={127,201,127}]{A}
    \Vertex[x=1,RGB,color={190,174,212}]{B}
    \Vertex[x=2,RGB,color={253,192,134}]{C}
  \end{tikzpicture}
\end{marginfigure}

\begin{docspec}
\begin{lstlisting}
  \begin{tikzpicture}
    \Vertex[RGB,color={127,201,127}]{A}
    \Vertex[x=1,RGB,color={190,174,212}]{B}
    \Vertex[x=2,RGB,color={253,192,134}]{C}
  \end{tikzpicture}
\end{lstlisting}
\end{docspec}

\begin{docspecdef}
  \doccmd{Vertex}[\docopt{Pseudo}]\{\docarg{Name}\}
\end{docspecdef}

The option \docopt{Pseudo} creates a pseudo vertex, where only the vertex name and the vertex coordinate will be drawn. Edges etc, can still be assigned to this vertex.

\begin{marginfigure}[32mm]
\centering
  \begin{tikzpicture}
    \Vertex{A}
    \Vertex[x=2,Pseudo]{B}
  \end{tikzpicture}
\end{marginfigure}

\begin{docspec}
\begin{lstlisting}
  \begin{tikzpicture}
    \Vertex{A}
    \Vertex[x=2,Pseudo]{B}
  \end{tikzpicture}
\end{lstlisting}
\end{docspec}

\begin{docspecdef}
  \doccmd{Vertex}[\docopt{layer}=\docarg{number}]\{\docarg{Name}\}
\end{docspecdef}

With the option \docopt{layer} the vertex can be assigned to a specific layer. More about this option and the use of layers is explained in Chapter~\ref{chap:multilayer_networks}.

\newpage

\section{Edge}
\label{sec:edge}

The second essential command is an \doccmddef{Edge}, which allow connecting two vertices.

\begin{docspecdef}
  \doccmd{Edge}[\docopt{local options}](\docarg{Vertex i})(\docarg{Vertex j})
\end{docspecdef}

Edges can be generated between one or two vertices. In the first case, a self-loop will be generated. As mandatory arguments the \docarg{Names} of the vertices which should be connected must be entered between $(~)$ brackets. In case of a directed edge, the order is important. An edge is created from \docarg{Vertex i} (origin) to \docarg{Vertex j} (destination).

For an \doccmd{Edge} the following options are available:

\begin{table}[h]\index{Edge!options}
  \footnotesize%
  \begin{center}
    \begin{tabular}{lccl}
      \toprule
      Option & Default & Type &Definition \\
      \midrule
      lw          & \{\}  & measure & line width of the edge \\
      color       & \{\}  & color   & edge color\\
      opacity     & \{\}  & number  & opacity of the edge \\
      bend        & 0     & number  & angle out/in of the vertex \\
      label       & \{\}  & string  & label \\
      fontsize    & \{\}  & string  & font size of the label\\
      fontcolor   & \{\}  & color   & font color of the label\\
      fontscale   & \{\}  & number  & scale of the label\\
      position    & \{\}  & string  & label position\\
      distance    & 0.5   & number  & label distance from vertex i\\
      style       & \{\}  & string  & additional \tikzsym styles \\
      \midrule
      path        & \{\}  & list    & path over several vertices\\
      \midrule
      loopsize    & 1cm   & measure & size parameter of the self-loop\\
      loopposition& 0     & number  & orientation of the self-loop \\
      loopshape   & 90    & number  & loop angle out/in of the vertex \\
      \midrule
      Direct      & false & Boolean & allow directed edges \\
      Math        & false & Boolean & displays the label in math mode \\
      RGB         & false & Boolean & allow RGB colors \\
      NotInBG     & false & Boolean & edge is not in the background layer\\
      \bottomrule
    \end{tabular}
    \scriptsize
  \end{center}
  \caption{Local options for the \doccmd{Edge} command.}
  \label{tab:edge_options}
\end{table}

The options \docopt{loopsize}, \docopt{loopposition}, and \docopt{loopsize} are only for self-loops available.

\begin{docspecdef}
  \doccmd{Edge}(\docarg{Vertex i})(\docarg{Vertex j})
\end{docspecdef}

An edge is created between \docarg{Vertex i} and \docarg{Vertex j}.

\begin{marginfigure}[28mm]
\centering
  \begin{tikzpicture}
    \Vertex{A} \Vertex[x=2]{B}
    \Edge(A)(B)
  \end{tikzpicture}
\end{marginfigure}

\begin{docspec}
\begin{lstlisting}
  \begin{tikzpicture}
    \Vertex{A} \Vertex[x=2]{B}
    \Edge(A)(B)
  \end{tikzpicture}
\end{lstlisting}
\end{docspec}

\begin{docspecdef}
  \doccmd{Edge}[\docopt{lw}=\docarg{measure}](\docarg{Vertex i})(\docarg{Vertex j})
\end{docspecdef}

The line width of an edge can be modified with the option \docopt{lw}. Here, the unit of the \docarg{measure} has to be specified. The default value is \SI{1.5}{pt}.

\newpage
\begin{marginfigure}[6mm]
\centering
  \begin{tikzpicture}
    \Vertex{A} \Vertex[x=2]{B} \Vertex[x=2,y=-1]{C}
    \Edge[lw=3pt](A)(B)
    \Edge[lw=5pt](A)(C)
  \end{tikzpicture}
\end{marginfigure}

\begin{docspec}
\begin{lstlisting}
  \begin{tikzpicture}
    \Vertex{A} \Vertex[x=2]{B} \Vertex[x=2,y=-1]{C}
    \Edge[lw=3pt](A)(B)
    \Edge[lw=5pt](A)(C)
  \end{tikzpicture}
\end{lstlisting}
\end{docspec}


\begin{docspecdef}
  \doccmd{Edge}[\docopt{color}=\docarg{color}](\docarg{Vertex i})(\docarg{Vertex j})
\end{docspecdef}

To change the line color of each edge individually, the option \docopt{color} has to be used. Without the option \docopt{RGB} set, the default \tikzsym and \LaTeX~ colors can be applied.

\begin{marginfigure}[30mm]
\centering
  \begin{tikzpicture}
    \Vertex{A} \Vertex[x=2]{B} \Vertex[x=2,y=-1]{C}
    \Edge[color=red](A)(B)
    \Edge[color=green!70!blue](A)(C)
  \end{tikzpicture}
\end{marginfigure}

\begin{docspec}
\begin{lstlisting}
  \begin{tikzpicture}
    \Vertex{A} \Vertex[x=2]{B} \Vertex[x=2,y=-1]{C}
    \Edge[color=red](A)(B)
    \Edge[color=green!70!blue](A)(C)
  \end{tikzpicture}
\end{lstlisting}
\end{docspec}

\begin{docspecdef}
  \doccmd{Edge}[\docopt{opacity}=\docarg{number}](\docarg{Vertex i})(\docarg{Vertex j})
\end{docspecdef}

With the option \docopt{opacity} the opacity of the edge line can be modified. The range of the \docarg{number} lies between $0$ and $1$. Where $0$ represents a fully transparent fill and $1$ a solid fill.

\begin{marginfigure}[30mm]
\centering
  \begin{tikzpicture}
    \Vertex{A} \Vertex[x=2]{B} \Vertex[x=2,y=-1]{C}
    \fill [orange] (.9,.2) rectangle (1.1,-.7);
    \EdgesNotInBG
    \Edge[opacity=.7](A)(B)
    \Edge[opacity=.2](A)(C)
  \end{tikzpicture}
\end{marginfigure}

\begin{docspec}
\begin{lstlisting}
  \begin{tikzpicture}
    \Vertex{A} \Vertex[x=2]{B} \Vertex[x=2,y=-1]{C}
    \Edge[opacity=.7](A)(B)
    \Edge[opacity=.2](A)(C)
  \end{tikzpicture}
\end{lstlisting}
\end{docspec}

\begin{docspecdef}
  \doccmd{Edge}[\docopt{bend}=\docarg{number}](\docarg{Vertex i})(\docarg{Vertex j})
\end{docspecdef}

The shape of the edge can be modified with the \docopt{bend} option. If nothing is specified a straight edge, between the vertices, is drawn. The \docarg{number} defines the angle in which the edge is diverging from its straight connection. A positive \docarg{number} bend the edge counter clockwise, while a negative \docarg{number} make changes clockwise.

\begin{marginfigure}[32mm]
\centering
  \begin{tikzpicture}
    \Vertex{A}
    \Vertex[x=2]{B}
    \Edge[bend=45](A)(B)
    \Edge[bend=-70](A)(B)
    \draw[orange,dashed](0,0) -- (20mm,0mm)(0,0)--(45:10mm);
    \draw[orange,->] (.5,0) arc (0:45:5mm);
    \draw[orange,dashed](0,0)--(-70:10mm);
    \draw[orange,->] (.5,0) arc (0:-70:5mm);
    \node[orange] (x) at (1,.25) {$45^\circ$};
    \node[orange] (x) at (1,-.35) {$70^\circ$};
    \draw[orange,dashed] (2,0)--++(135:10mm);
    \draw[orange,->] (1.5,0) arc (180:135:5mm);
    \draw[orange,dashed] (2,0)--++(-110:10mm);
    \draw[orange,->] (1.5,0) arc (-180:-110:5mm);
  \end{tikzpicture}
\end{marginfigure}

\begin{docspec}
\begin{lstlisting}
  \begin{tikzpicture}
    \Vertex{A} \Vertex[x=2]{B}
    \Edge[bend=45](A)(B)
    \Edge[bend=-70](A)(B)
  \end{tikzpicture}
\end{lstlisting}
\end{docspec}

\begin{docspecdef}
  \doccmd{Edge}[\docopt{label}=\docarg{string}](\docarg{Vertex i})(\docarg{Vertex j})
\end{docspecdef}

An edge is labeled with the option \docopt{label}. For the label any \docarg{string} argument can be used, including blank spaces. The environment \$ \$ can be used to display mathematical expressions.

\begin{marginfigure}[30mm]
\centering
  \begin{tikzpicture}
    \Vertex{A} \Vertex[x=2]{B}
    \Edge[label=X](A)(B)
  \end{tikzpicture}
\end{marginfigure}

\begin{docspec}
\begin{lstlisting}
  \begin{tikzpicture}
    \Vertex{A} \Vertex[x=2]{B}
    \Edge[label=X](A)(B)
  \end{tikzpicture}
\end{lstlisting}
\end{docspec}

\begin{docspecdef}
  \doccmd{Edge}[\docopt{label}=\docarg{string},\docopt{fontsize}=\docarg{string}](\docarg{Vertex i})(\docarg{Vertex j})
\end{docspecdef}

The font size of the \docopt{label} can be modified with the option \docopt{fontsize}. Here common \LaTeX~ font size commands\footnote{e.g. \doccmd{tiny}, \doccmd{scriptsize}, \doccmd{footnotesize}, \doccmd{small}, \dots.} can be used to change the size of the label.

\begin{marginfigure}[6mm]
\centering
  \begin{tikzpicture}
    \Vertex{A} \Vertex[x=2]{B} \Vertex[x=2,y=-1]{C}
    \Edge[label=X,fontsize=\large](A)(B)
    \Edge[label=Y,fontsize=\tiny](A)(C)
  \end{tikzpicture}
\end{marginfigure}

\begin{docspec}
\begin{lstlisting}
  \begin{tikzpicture}
    \Vertex{A} \Vertex[x=2]{B} \Vertex[x=2,y=-1]{C}
    \Edge[label=X,fontsize=\large](A)(B)
    \Edge[label=Y,fontsize=\tiny](A)(C)
  \end{tikzpicture}
\end{lstlisting}
\end{docspec}

\begin{docspecdef}
  \doccmd{Edge}[\docopt{label}=\docarg{string},\docopt{fontcolor}=\docarg{color}](\docarg{Vertex i})(\docarg{Vertex j})
\end{docspecdef}

The color of the \docopt{label} can be changed with the option \docopt{fontcolor}. Currently, only the default \tikzsym and \LaTeX~ colors are supported \footnote{\TODO Add RGB option!}.

\begin{marginfigure}[1mm]
\centering
  \begin{tikzpicture}
    \Vertex{A} \Vertex[x=2]{B} \Vertex[x=2,y=-1]{C}
    \Edge[label=X,fontcolor=blue](A)(B)
    \Edge[label=Y,fontcolor=red](A)(C)
  \end{tikzpicture}
\end{marginfigure}

\begin{docspec}
\begin{lstlisting}
  \begin{tikzpicture}
    \Vertex{A} \Vertex[x=2]{B} \Vertex[x=2,y=-1]{C}
    \Edge[label=X,fontcolor=blue](A)(B)
    \Edge[label=Y,fontcolor=red](A)(C)
  \end{tikzpicture}
\end{lstlisting}
\end{docspec}

\begin{docspecdef}
  \doccmd{Edge}[\docopt{label}=\docarg{string},\docopt{fontscale}=\docarg{color}](\docarg{Vertex i})(\docarg{Vertex j})
\end{docspecdef}

Contrary to the option \docopt{fontsize}, the option \docopt{fontscale} does not change the font size itself but scales the curent font size up or down. The \docarg{number} defines the scale, where numbers between $0$ and $1$ down scale the font and numbers greater then $1$ up scale the label. For example $0.5$ reduces the size of the font to $50\%$ of its originial size, while $1.2$ scales the font to $120\%$.

\begin{marginfigure}[30mm]
\centering
  \begin{tikzpicture}
    \Vertex{A} \Vertex[x=2]{B} \Vertex[x=2,y=-1]{C}
    \Edge[label=X,fontscale=.5](A)(B)
    \Edge[label=Y,fontscale=1.2](A)(C)
  \end{tikzpicture}
\end{marginfigure}

\begin{docspec}
\begin{lstlisting}
  \begin{tikzpicture}
    \Vertex{A} \Vertex[x=2]{B} \Vertex[x=2,y=-1]{C}
    \Edge[label=X,fontscale=.5](A)(B)
    \Edge[label=Y,fontscale=2](A)(C)
  \end{tikzpicture}
\end{lstlisting}
\end{docspec}

\begin{docspecdef}
  \doccmd{Edge}[\docopt{label}=\docarg{string},\docopt{position}=\docarg{string}](\docarg{Vertex i})(\docarg{Vertex j})
\end{docspecdef}

Per default the \docopt{label} is positioned in between both vertices in the center of the line. Classical \tikzsym commands\footnote{e.g. \docarg{above}, \docarg{below}, \docarg{left}, \docarg{right}, \docarg{above left}, \docarg{above right},\dots} can be used to change the \docopt{position} of the \docopt{label}.

\begin{marginfigure}[6mm]
\centering
  \begin{tikzpicture}
    \Vertex{A} \Vertex[x=2]{B} \Vertex[x=2,y=-1]{C}
    \Edge[label=X,position=above](A)(B)
    \Edge[label=Y,position={below left=2mm}](A)(C)
  \end{tikzpicture}
\end{marginfigure}

\begin{docspec}
\begin{lstlisting}
  \begin{tikzpicture}
    \Vertex{A} \Vertex[x=2]{B} \Vertex[x=2,y=-1]{C}
    \Edge[label=X,position=above](A)(B)
    \Edge[label=Y,position={below left=2mm}](A)(C)
  \end{tikzpicture}
\end{lstlisting}
\end{docspec}

\begin{docspecdef}
  \doccmd{Edge}[\docopt{label}=\docarg{string},\docopt{distance}=\docarg{number}](\docarg{Vertex i})(\docarg{Vertex j})
\end{docspecdef}

The label position between the vertices can be modified with the \docopt{distance} option. Per default the \docopt{label} is centered between both vertices. The position is expressed as the percentage of the length between the vertices, e.g. of \docopt{distance}=$0.7$, the label is placed at 70\% of the edge length away of \docarg{Vertex i}.

\newpage
\begin{marginfigure}[6mm]
\centering
  \begin{tikzpicture}
    \Vertex{A} \Vertex[x=2]{B}
    \Edge[label=X,distance=.7](A)(B)
    \draw[orange,|->|] (.3,-.5) --++ (0.98,0) node[pos=.5,above]{$0.7$};
    \draw[orange,|<->|] (.3,.5) --++ (1.4,0) node[pos=.5,below]{$1.0$};
  \end{tikzpicture}
\end{marginfigure}

\begin{docspec}
\begin{lstlisting}
  \begin{tikzpicture}
    \Vertex{A} \Vertex[x=2]{B}
    \Edge[label=X,distance=.7](A)(B)
  \end{tikzpicture}
\end{lstlisting}
\end{docspec}

\begin{docspecdef}
  \doccmd{Edge}[\docopt{style}=\docarg{string}](\docarg{Vertex i})(\docarg{Vertex j})
\end{docspecdef}

Any other \tikzsym style option or command can be entered via the option \docopt{style}. Most of these commands can be found in the ``\tikzsym and PGF Manual''. Contain the commands additional options (e.g.\docopt{shading}=\docarg{ball}), then the argument for the \docopt{style} has to be between $\{~\}$ brackets.

\begin{marginfigure}[30mm]
\centering
  \begin{tikzpicture}
    \Vertex{A} \Vertex[x=2]{B}
    \Edge[style={dashed}](A)(B)
  \end{tikzpicture}
\end{marginfigure}

\begin{docspec}
\begin{lstlisting}
  \begin{tikzpicture}
    \Vertex{A} \Vertex[x=2]{B}
    \Edge[style={dashed}](A)(B)
  \end{tikzpicture}
\end{lstlisting}
\end{docspec}

\begin{docspecdef}
  \doccmd{Edge}[\docopt{path}=\docarg{list}](\docarg{Vertex i})(\docarg{Vertex j})
\end{docspecdef}

In order to draw a finite sequence of edges which connect a sequence of vertices and/or coordinates, the option \docopt{path} can be used\footnote{\TODO currently labels and bend edges are not supported!}. The argument for this option has to be a list element indicated by $\{~\}$ brackets, containing the \docarg{Names} of the intermediated vertices. New coordinates, i.e. there is no vertex located, can be insert with $\{\docopt{x},\docopt{y}\}$. Arguments of the list, have to be seperated by commas <<,>>.

\begin{marginfigure}[18mm]
\centering
  \begin{tikzpicture}
    \Vertex{A} \Vertex[x=2]{B} \Vertex[x=2,y=-1]{C}
    \Edge[path={A,{0,-1},C,B}](A)(B)
  \end{tikzpicture}
\end{marginfigure}

\begin{docspec}
\begin{lstlisting}
  \begin{tikzpicture}
    \Vertex{A} \Vertex[x=2]{B} \Vertex[x=2,y=-1]{C}
    \Edge[path={A,{0,-1},C,B}](A)(B)
  \end{tikzpicture}
\end{lstlisting}
\end{docspec}


\begin{docspecdef}
  \doccmd{Edge}(\docarg{Vertex i})(\docarg{Vertex i})
\end{docspecdef}

Self-loops are created by using the same vertex as origin and destination. Beside the options explained above, there are three self-loop specific options: \docopt{loopsize}, \docopt{loopposition}, and \docopt{loopshape}.

\begin{marginfigure}[28mm]
\centering
  \begin{tikzpicture}
    \Vertex{A}
    \Edge(A)(A)
  \end{tikzpicture}
\end{marginfigure}

\begin{docspec}
\begin{lstlisting}
  \begin{tikzpicture}
    \Vertex{A}
    \Edge(A)(A)
  \end{tikzpicture}
\end{lstlisting}
\end{docspec}

\begin{docspecdef}
  \doccmd{Edge}[\docopt{loopsize}=\docarg{measure}](\docarg{Vertex i})(\docarg{Vertex i})
\end{docspecdef}

With the option \docopt{loopsize} the length of the edge can be modified. The \docarg{measure} value has to be insert together with its units. Per default the \docopt{loopsize} is \SI{1}{cm}.

\newpage
\begin{marginfigure}[2mm]
\centering
  \begin{tikzpicture}
    \Vertex{A} \Vertex[x=1.3]{B}
    \Edge[loopsize=.5cm](A)(A)
    \Edge[loopsize=1.5cm](B)(B)
  \end{tikzpicture}
\end{marginfigure}

\begin{docspec}
\begin{lstlisting}
  \begin{tikzpicture}
    \Vertex{A} \Vertex[x=1.3]{B}
    \Edge[loopsize=.5cm](A)(A)
    \Edge[loopsize=1.5cm](B)(B)
  \end{tikzpicture}
\end{lstlisting}
\end{docspec}

\begin{docspecdef}
  \doccmd{Edge}[\docopt{loopposition}=\docarg{number}](\docarg{Vertex i})(\docarg{Vertex i})
\end{docspecdef}

The position of the self-loop is defined via the rotation angle around the vertex. The origin ($0^\circ$) is the $y$ axis. A positive \docarg{number} change the \docopt{loopposition} counter clockwise, while a negative \docarg{number} make changes clockwise.

\begin{marginfigure}[28mm]
\centering
  \begin{tikzpicture}
    \Vertex{A}
    \Vertex[x=1.5]{B}
    \Edge[loopposition=45](A)(A)
    \Edge[loopposition=-70](B)(B)
    \draw[orange,dashed](0,0) -- (10mm,0mm)(0,0)--(45:10mm);
    \draw[orange,->] (.5,0) arc (0:45:5mm);
    \node[orange] (x) at (1,.35) {$45^\circ$};
    \draw[orange,dashed](1.5,0)--++(-70:10mm);
    \draw[orange,dashed](1.5,0)--++(10mm,0);
    \draw[orange,->] (1.9,0) arc (0:-70:5mm);
    \node[orange] (x) at (2.3,-.45) {$70^\circ$};
  \end{tikzpicture}
\end{marginfigure}

\begin{docspec}
\begin{lstlisting}
  \begin{tikzpicture}
    \Vertex{A} \Vertex[x=1.5]{B}
    \Edge[loopposition=45](A)(A)
    \Edge[loopposition=-70](B)(B)
  \end{tikzpicture}
\end{lstlisting}
\end{docspec}

\begin{docspecdef}
  \doccmd{Edge}[\docopt{loopshape}=\docarg{number}](\docarg{Vertex i})(\docarg{Vertex i})
\end{docspecdef}

The shape of the self-loop is defined by the enclosing angle. The shape can be changed by decreasing or increasing the argument value of the \docopt{loopshape} option.

\begin{marginfigure}[28mm]
\centering
  \begin{tikzpicture}
    \Vertex{A}
    \Edge[loopshape=45](A)(A)
    \draw[orange,dashed](0,0) -- (-22.5:13mm)(0,0)--(22.5:13mm);
    \draw[orange,->] (1.2,0) arc (0:22.5:12mm);
    \draw[orange,->] (1.2,0) arc (0:-22.5:12mm);
    \node[orange] (x) at (1.5,0) {$45^\circ$};
  \end{tikzpicture}
\end{marginfigure}

\begin{docspec}
\begin{lstlisting}
  \begin{tikzpicture}
    \Vertex{A}
    \Edge[loopshape=45](A)(A)
  \end{tikzpicture}
\end{lstlisting}
\end{docspec}


\begin{docspecdef}
  \doccmd{Edge}[\docopt{Direct}](\docarg{Vertex i})(\docarg{Vertex j})
\end{docspecdef}

Directed edges are created by enabling the option \docopt{Direct}. The arrow is drawn from \docarg{Vertex i} to \docarg{Vertex j}.

\begin{marginfigure}[28mm]
\centering
  \begin{tikzpicture}
    \Vertex{A} \Vertex[x=2]{B}
    \Edge[Direct](A)(B)
  \end{tikzpicture}
\end{marginfigure}

\begin{docspec}
\begin{lstlisting}
  \begin{tikzpicture}
    \Vertex{A} \Vertex[x=2]{B}
    \Edge[Direct](A)(B)
  \end{tikzpicture}
\end{lstlisting}
\end{docspec}

\begin{docspecdef}
  \doccmd{Edge}[Math, label=\docopt{string}](\docarg{Vertex i})(\docarg{Vertex j})
\end{docspecdef}

The option \docopt{Math} allows transforming labels into mathematical expressions without using the \$~\$ environment.

\begin{marginfigure}[28mm]
\centering
  \begin{tikzpicture}
    \Vertex{A} \Vertex[x=2]{B}
    \Edge[Math,label=X_1](A)(B)
  \end{tikzpicture}
\end{marginfigure}

\begin{docspec}
\begin{lstlisting}
  \begin{tikzpicture}
    \Vertex{A} \Vertex[x=2]{B}
    \Edge[Math,label=X_1](A)(B)
  \end{tikzpicture}
\end{lstlisting}
\end{docspec}

\begin{docspecdef}
  \doccmd{Edge}[RBG, color=\docopt{RGB value}](\docarg{Vertex i})(\docarg{Vertex j})
\end{docspecdef}

In order to display RGB colors for the line color of the edge, the option \docopt{RGB} has to be entered. In combination with this option, the \docopt{color} hast to be a list with the \docarg{RGB
values}, separated by <<,>> and within $\{~\}$.\footnote{e.g. the RGB code for white: $\{255,255,255\}$}

\begin{marginfigure}
\centering
  \begin{tikzpicture}
    \Vertex{A} \Vertex[x=2]{B} \Vertex[x=2,y=-1]{C}
    \Edge[RGB,color={127,201,127}](A)(B)
    \Edge[RGB,color={253,192,134}](A)(C)
  \end{tikzpicture}
\end{marginfigure}

\begin{docspec}
\begin{lstlisting}
  \begin{tikzpicture}
    \Vertex{A} \Vertex[x=2]{B} \Vertex[x=2,y=-1]{C}
    \Edge[RGB,color={127,201,127}](A)(B)
    \Edge[RGB,color={253,192,134}](A)(C)
  \end{tikzpicture}
\end{lstlisting}
\end{docspec}

\begin{docspecdef}
  \doccmd{Edge}[\docopt{NotInBG}]\{\docarg{filename}\}
\end{docspecdef}

Per default, the edge is drawn on the background layer of the \docarg{tikzpicture}. I.e. objects which are created after the edges appear also on top of them. To turn this off, the option \docopt{NotInBG} has to be enabled. Changes to the default setting can be made with \doccmd{EdgesNotInBG} or \doccmd{EdgesInBG}\footnote{See Section \ref{sec:edge_style}}.

\begin{marginfigure}[6mm]
\centering
  \begin{tikzpicture}
    \Vertex{A} \Vertex[x=2]{B} \Vertex[x=1,y=-.5]{C}
    \Vertex[y=-1]{D} \Vertex[x=2,y=-1]{E}
    \Edge[bend=-30](A)(B)
    \Edge[bend=30,NotInBG](D)(E)
  \end{tikzpicture}
\end{marginfigure}

\begin{docspec}
\begin{lstlisting}
  \begin{tikzpicture}
    \Vertex{A} \Vertex[x=2]{B} \Vertex[x=1,y=-.5]{C}
    \Vertex[y=-1]{D} \Vertex[x=2,y=-1]{E}
    \Edge[bend=-30](A)(B)
    \Edge[bend=30,NotInBG](D)(E)
  \end{tikzpicture}
\end{lstlisting}
\end{docspec}

\newpage
\section{Text}
\label{sec:text}

While \tikzsym offers multiple ways to label objects and create text elements, a simplified command \doccmddef{Text} is implemented, which allow placing and modifying text to the networks.
\begin{docspecdef}
  \doccmd{Text}[\docopt{local options}]\{\docarg{string}\}
\end{docspecdef}
In order to be able to create a text, a non-empty \docarg{string} argument is required. This argument is the actual text added to the figure. Mathematical symbols are entered in the same way as in a normal \LaTeX~document, i.e. between \$~\$.

For a \doccmd{Text} the following options are available:

\begin{table}[h]\index{Text!options}
  \footnotesize%
  \begin{center}
    \begin{tabular}{lccl}
      \toprule
      Option & Default & Type &Definition \\
      \midrule
      x          & 0     & measure  & x-coordinate \\
      y          & 0     & measure  & y-coordinate \\
      fontsize   & \{\}  & fontsize & font size of the text\\
      color      & \{\}  & color    & color of the text\\
      opacity    & \{\}  & number   & opacity of the text \\
      position   & center& string   & position of the text to the origin\\
      distance   & \SI{0}{cm} & measure  & distance from the origin \\
      rotation   & 0     & number   & rotation of the text \\
      anchor     & \{\}  & string   & anchor of the text \\
      width      & \{\}  & number   & width of the text box \\
      style      & \{\}  & string   & additional \tikzsym styles \\
      layer      & \{\}  & number   & assigned layer of the text \\
      \midrule
      RGB        & false & Boolean  & allow RGB colors \\
      \bottomrule
    \end{tabular}
    \scriptsize
  \end{center}
  \caption{Local options for the \doccmd{Text} command.}
  \label{tab:text_options}
\end{table}

The order how the options are entered does not matter. Changes to the default Text layout can be made with \doccmd{SetTextStyle}\footnote{see Section \ref{sec:text_style}}

\begin{docspecdef}
  \doccmd{Text}[\docopt{x}=\docarg{measure},\docopt{y}=\docarg{measure}]\{\docarg{string}\}
\end{docspecdef}

The location of the text is determined by Cartesian coordinates in \docopt{x} and \docopt{y}. The coordinates are optional. If no coordinates are determined the text will be placed at the origin ($0,0$). The entered \docarg{measures} are in default units (\si{cm}). Changing the unites (locally) can be done by adding the unit to the \docarg{measure}\footnote{e.g. x=\SI{1}{in}}. Changes to the default setting can be made with \doccmd{SetDefaultUnit}\footnote{see Section \ref{sec:gerneral_settings}}.

\begin{marginfigure}[6mm]
\centering
  \begin{tikzpicture}
    \Text{A}
    \Text[x=1,y=1]{B}
    \Text[x=2]{C}
  \end{tikzpicture}
\end{marginfigure}

\begin{docspec}
\begin{lstlisting}
  \begin{tikzpicture}
    \Text{A}
    \Text[x=1,y=1]{B}
    \Text[x=2]{C}
  \end{tikzpicture}
\end{lstlisting}
\end{docspec}

\begin{docspecdef}
  \doccmd{Text}[\docopt{fontsize}=\docarg{font size}]\{\docarg{string}\}
\end{docspecdef}

The font size of the text can be changed with the option \docopt{fontsize}. Per default the font size of the text is defined as \doccmd{normalsize}.

\begin{marginfigure}[34mm]
\centering
  \begin{tikzpicture}
    \Text[fontsize=\small]{A}
    \Text[x=1,fontsize=\LARGE]{B}
    \Text[x=2,fontsize=\Huge]{C}
  \end{tikzpicture}
\end{marginfigure}

\begin{docspec}
\begin{lstlisting}
  \begin{tikzpicture}
    \Text[fontsize=\small]{A}
    \Text[x=1,fontsize=\LARGE]{B}
    \Text[x=2,fontsize=\Huge]{C}
  \end{tikzpicture}
\end{lstlisting}
\end{docspec}

\begin{docspecdef}
  \doccmd{Text}[\docopt{color}=\docarg{color}]\{\docarg{string}\}
\end{docspecdef}

To change the text color individually, the option \docopt{color} has to be used. Without the option \docopt{RGB} set, the default \tikzsym and \LaTeX~ colors can be applied.

\begin{marginfigure}[34mm]
\centering
  \begin{tikzpicture}
    \Text[color = blue]{A}
    \Text[x=1,color=red]{B}
    \Text[x=2,color=green!70!blue]{C}
  \end{tikzpicture}
\end{marginfigure}

\begin{docspec}
\begin{lstlisting}
  \begin{tikzpicture}
    \Text[color = blue]{A}
    \Text[x=1,color=red]{B}
    \Text[x=2,color=green!70!blue]{C}
  \end{tikzpicture}
\end{lstlisting}
\end{docspec}

\begin{docspecdef}
  \doccmd{Text}[\docopt{opacity}=\docarg{number}]\{\docarg{string}\}
\end{docspecdef}

With the option \docopt{opacity} the opacity of the text can be modified. The range of the \docarg{number} lies between $0$ and $1$. Where $0$ represents a fully transparent text and $1$ a solid text.

\begin{marginfigure}[34mm]
\centering
  \begin{tikzpicture}
    \draw[thick,orange,dashed] (-.5,0) -- (2.5,0);
    \Text[opacity = 1]{A}
    \Text[x=1,opacity =.7]{B}
    \Text[x=2,opacity =.2]{C}
  \end{tikzpicture}
\end{marginfigure}

\begin{docspec}
\begin{lstlisting}
  \begin{tikzpicture}
    \Text[opacity = 1]{A}
    \Text[x=1,opacity =.7]{B}
    \Text[x=2,opacity =.2]{C}
  \end{tikzpicture}
\end{lstlisting}
\end{docspec}

\begin{docspecdef}
 \doccmd{Text}[\docopt{position}=\docarg{string},\docopt{distance}=\docarg{measure}]\{\docarg{string}\}
\end{docspecdef}

Per default the \docopt{position} of the text is in the \docarg{center} of the origin. Classical \tikzsym commands\footnote{e.g. \docarg{above}, \docarg{below}, \docarg{left}, \docarg{right}, \docarg{above left}, \docarg{above right},\dots} can be used to change the \docopt{position} of the text.

With the option, \docopt{distance} the distance between the text and the origin can be changed.

\begin{marginfigure}[20mm]
\centering
  \begin{tikzpicture}
    \filldraw [orange] (0,0) circle (2pt);
    \draw[<-,orange] (0.1,0) --++ (0.5,0) node [pos=1,right]{origin ($0,0$)};
    \Text[position=above]{above}
    \Text[position=below]{below}
    \Text[position=left,distance=5mm]{left}
    \Text[position=above right,distance=5mm]{above right}
  \end{tikzpicture}
\end{marginfigure}

\begin{docspec}
\begin{lstlisting}
  \begin{tikzpicture}
    \Text[position=above]{above}
    \Text[position=below]{below}
    \Text[position=left,distance=5mm]{left}
    \Text[position=above right,distance=5mm]{above right}
  \end{tikzpicture}
\end{lstlisting}
\end{docspec}

\begin{docspecdef}
 \doccmd{Text}[\docopt{rotation}=\docarg{number}]\{\docarg{string}\}
\end{docspecdef}

With the \docopt{rotation}, the text can be rotated by entering a \docarg{number} between $-360$ and $360$. The origin ($0^\circ$) is the $y$ axis. A positive \docarg{number} change the \docopt{position} counter clockwise, while a negative \docarg{number} make changes clockwise.

\begin{marginfigure}[31mm]
\centering
  \begin{tikzpicture}
    \Text[rotation = 30]{A}
    \Text[x=1,rotation = 45]{B}
    \Text[x=2,rotation = 75]{C}
    \draw[orange,dashed](2,0) --++ (7mm,0mm)(2,0)--++(75:7mm);
    \draw[orange,->] (2.5,0) arc (0:75:5mm);
    \node[orange] (x) at (2.6,-.2) {$75^\circ$};
  \end{tikzpicture}
\end{marginfigure}

\begin{docspec}
\begin{lstlisting}
  \begin{tikzpicture}
    \Text[rotation=30]{A}
    \Text[x=1,rotation=45]{B}
    \Text[x=2,rotation=75]{C}
  \end{tikzpicture}
\end{lstlisting}
\end{docspec}
\newpage

\begin{docspecdef}
 \doccmd{Text}[\docopt{anchor}=\docarg{string}]\{\docarg{string}\}
\end{docspecdef}

With the option \docopt{anchor} the alignment of the text can be changed. Per default the text will be aligned centered. Classical \tikzsym commands\footnote{e.g. \docarg{north}, \docarg{east}, \docarg{south}, \docarg{west}, \docarg{north east}, \docarg{north west},\dots} can be used to change the alignment of the text.

\begin{marginfigure}[5mm]
\centering
  \begin{tikzpicture}
    \filldraw [orange] (0,0) circle (2pt);
    \filldraw [orange] (1,0) circle (2pt);
    \filldraw [orange] (2,0) circle (2pt);
    \Text[anchor=north east]{NE}
    \Text[x=1,anchor = south]{S}
    \Text[x=2,anchor =south west]{SW}
  \end{tikzpicture}
\end{marginfigure}

\begin{docspec}
\begin{lstlisting}
  \begin{tikzpicture}
    \Text[anchor=north east]{NE}
    \Text[x=1,anchor = south]{S}
    \Text[x=2,anchor =south west]{SW}
  \end{tikzpicture}
\end{lstlisting}
\end{docspec}

\begin{docspecdef}
 \doccmd{Text}[\docopt{width}=\docarg{measure}]\{\docarg{string}\}
\end{docspecdef}

With the option \docopt{width} enabled, the text will break after the entered \docarg{measure}.

\begin{marginfigure}[25mm]
\centering
  \begin{tikzpicture}
    \draw[|<->|,orange] (-1.25,.7) --++ (2.5,0) node [pos=.5,above]{2.5cm};
    \Text[width=2.5cm]{This might be a very long text.}
  \end{tikzpicture}
\end{marginfigure}

\begin{docspec}
\begin{lstlisting}
  \begin{tikzpicture}
    \Text[width=2.5cm]{This might be a very long text.}
  \end{tikzpicture}
\end{lstlisting}
\end{docspec}

\begin{docspecdef}
  \doccmd{Text}[\docopt{style}=\{\docarg{string}\}]\{\docarg{string}\}
\end{docspecdef}

Any other \tikzsym style option or command can be entered via the option \docopt{style}. Most of these commands can be found in the ``\tikzsym and PGF Manual''. Contain the commands additional options (e.g.\docopt{fill}=\docarg{red}), then the argument for the \docopt{style} has to be between $\{~\}$ brackets.

\begin{marginfigure}[32mm]
\centering
  \begin{tikzpicture}
    \Text[style={draw,rectangle}]{A}
    \Text[x=1,style={fill=red}]{B}
    \Text[x=2,style={fill=blue,circle,opacity=.3}]{C}
  \end{tikzpicture}
\end{marginfigure}

\begin{docspec}
\begin{lstlisting}
  \begin{tikzpicture}
    \Text[style={draw,rectangle}]{A}
    \Text[x=1,style={fill=red}]{B}
    \Text[x=2,style={fill=blue,circle,opacity=.3}]{C}
  \end{tikzpicture}
\end{lstlisting}
\end{docspec}

\begin{docspecdef}
  \doccmd{Text}[\docopt{RGB},\docopt{color}=\docarg{RGB values}]\{\docarg{string}\}
\end{docspecdef}

In order to display RGB colors for the text color, the option \docopt{RGB} has to be entered. In combination with this option, the \docopt{color} hast to be a list with the \docarg{RGB values}, separated by <<,>> and within $\{~\}$.\footnote{e.g. the RGB code for white: $\{255,255,255\}$}

\begin{marginfigure}[7mm]
\centering
  \begin{tikzpicture}
    \Text[RGB,color={127,201,127}]{A}
    \Text[x=1,RGB,color={190,174,212}]{B}
    \Text[x=2,RGB,color={253,192,134}]{C}
  \end{tikzpicture}
\end{marginfigure}

\begin{docspec}
\begin{lstlisting}
  \begin{tikzpicture}
    \Text[RGB,color={127,201,127}]{A}
    \Text[x=1,RGB,color={190,174,212}]{B}
    \Text[x=2,RGB,color={253,192,134}]{C}
  \end{tikzpicture}
\end{lstlisting}
\end{docspec}

\begin{docspecdef}
  \doccmd{layer}[\docopt{layer}=\docarg{number}]\{\docarg{string}\}
\end{docspecdef}

With the option \docopt{layer} the text can be assigned to a specific layer. More about this option and the use of layers is explained in Chapter~\ref{chap:multilayer_networks}.

\chapter{Complex Networks}
\label{chap:complex_networks}

While in Chapter~\ref{chap:simple_networks} the building blocks of the networks are introduced, here the main strength of the \pkg package is explained. This includes creating networks based on data, obtained from other sources (e.g. Python, R, GIS). The idea is that the layout will be done by this external sources and \pkg is used make some changes and to recreate the networks in \LaTeX.

\section{Vertices}
\label{sec:vertices}

The \doccmddef{Vertices} command is the extension of the \doccmd{Vertex} command. Instead of a single vertex, a set of vertices will be drawn. This set of vertices is defined in an external file but can be modified with \doccmd{Vertices}.
\begin{docspecdef}
  \doccmd{Vertices}[\docopt{global options}]\{\docarg{filename}\}
\end{docspecdef}

The vertices have to be stored in a clear text file\footnote{e.g. .txt, .tex, .csv, .dat, \dots}, preferentially in a \texttt{.csv} format. The first row should contain the headings, which are equal to the options defined in Table \ref{tab:vertex_options}. Option are separated by a comma <<,>>. Each new row is corresponds to a new vertex.
\marginnote[8mm]{File: vertices.csv}
\begin{docspec}
\begin{lstlisting}
id, x, y ,size,color ,opacity,label,IdAsLabel,NoLabel
 A, 0,  0, .4 ,green ,  .9   ,  a  ,  false  , false
 B, 1, .7, .6 ,      ,  .5   ,  b  ,  false  , false
 C, 2,  1, .8 ,orange,  .3   ,  c  ,  false  , true
 D, 2,  0, .5 ,red   ,  .7   ,  d  ,  true   , false
 E,.2,1.5, .5 ,gray  ,       ,  e  ,  false  , false
\end{lstlisting}
\end{docspec}

Only the \docopt{id} value is mandatory for a vertex and corresponds to the \docarg{Name} argument of a single \doccmd{Vertex}. Therefore, the same rules and naming conventions apply as for the \docarg{Name} argument: no mathematical expressions, no blank spaces, and the \docopt{id} must be unique! All other options are optional. No specific order of the options must be maintained. If no value is entered for an option, the default value will be chosen\footnote{\TODO This is NOT valid for Boolean options, here values for all vertices have to be entered.}. The \docarg{filename} should not contain blank spaces or special characters. The vertices are drawn by the command \doccmd{Vertex} with the \docarg{filename} plus file format (e.g. \texttt{.csv}). If the vertices file is not in the same directory as the main \LaTeX~file, also the path has to be specified.

\newpage
\begin{marginfigure}[2mm]
\centering
  \begin{tikzpicture}
    \Vertices{data/vertices.csv}
  \end{tikzpicture}
\end{marginfigure}

\begin{docspec}
\begin{lstlisting}
  \begin{tikzpicture}
    \Vertices{data/vertices.csv}
  \end{tikzpicture}
\end{lstlisting}
\end{docspec}

Predefined \doccmd{Vertex} options can be overruled by the \docopt{global options} of the \doccmd{Vertices} command; I.e. these options apply for all vertices in the file. For the \doccmd{Vertices} the following options are available:

\begin{table}[h]\index{Vertices!options}
  \footnotesize%
  \begin{center}
    \begin{tabular}{lccl}
      \toprule
      Option & Default & Type &Definition \\
      \midrule
      size       & \{\}  & measure& diameter of the circles\\
      color      & \{\}  & color & fillcolor of vertices\\
      opacity    & \{\}  & number & opacity of the fill color \\
      style      & \{\}  & string & additional \tikzsym styles \\
      layer      & \{\}  & number & assigned layer of the vertices \\
      \midrule
      NoLabel    & false & Boolean & delete the labels \\
      IdAsLabel  & false & Boolean & uses the \docarg{Names} as labels \\
      Math       & false & Boolean & displays the labels in math mode \\
      RGB        & false & Boolean & allow RGB colors \\
      Pseudo     & false & Boolean & create a pseudo vertices\\
      \bottomrule
    \end{tabular}
    \scriptsize
  \end{center}
  \caption{Global options for the \doccmd{Vertices} command.}
  \label{tab:vertices_options}
\end{table}

The use of these options are similar to the options for a single \doccmd{Vertex} defined in Section~\ref{sec:vertex}.

\begin{docspecdef}
  \doccmd{Vertices}[\docopt{size}=\docarg{measure}]\{\docarg{filename}\}
\end{docspecdef}

The diameter of the vertices can be changed with the option \docopt{size}. Per default a vertex has \SI{0.6}{cm} in diameter. Also, here the default units are \si{cm} and have not to be added to the \docarg{measure}.

\begin{marginfigure}[28mm]
\centering
  \begin{tikzpicture}
    \Vertices[size=.6]{data/vertices.csv}
  \end{tikzpicture}
\end{marginfigure}

\begin{docspec}
\begin{lstlisting}
  \begin{tikzpicture}
    \Vertices[size=.6]{data/vertices.csv}
  \end{tikzpicture}
\end{lstlisting}
\end{docspec}

\begin{docspecdef}
  \doccmd{Vertices}[\docopt{color}=\docarg{color}]\{\docarg{filename}\}
\end{docspecdef}

To change the fill color for all vertices, the option \docopt{color} has to be used. Without the option \docopt{RGB} set, the default \tikzsym and \LaTeX~ colors can be applied.

\begin{marginfigure}[22mm]
\centering
  \begin{tikzpicture}
    \Vertices[color=green!70!blue]{data/vertices.csv}
  \end{tikzpicture}
\end{marginfigure}

\begin{docspec}
\begin{lstlisting}
  \begin{tikzpicture}
    \Vertices[color=green!70!blue]{data/vertices.csv}
  \end{tikzpicture}
\end{lstlisting}
\end{docspec}

\begin{docspecdef}
  \doccmd{Vertices}[\docopt{opacity}=\docarg{number}]\{\docarg{filename}\}
\end{docspecdef}

With the option \docopt{opacity} the opacity of all vertices fills colors can be modified. The range of the \docarg{number} lies between $0$ and $1$. Where $0$ represents a fully transparent fill and $1$ a solid fill.

\newpage
\begin{marginfigure}[2mm]
\centering
  \begin{tikzpicture}
    \Vertices[opacity=.3]{data/vertices.csv}
  \end{tikzpicture}
\end{marginfigure}

\begin{docspec}
\begin{lstlisting}
  \begin{tikzpicture}
    \Vertices[opacity=.3]{data/vertices.csv}
  \end{tikzpicture}
\end{lstlisting}
\end{docspec}

\begin{docspecdef}
  \doccmd{Vertices}[\docopt{style}=\docarg{string}]\{\docarg{filename}\}
\end{docspecdef}

Any other \tikzsym style option or command can be entered via the option \docopt{style}. Most of these commands can be found in the ``\tikzsym and PGF Manual''. Contain the commands additional options (e.g.\docopt{shading}=\docarg{ball}), then the argument for the \docopt{style} has to be between $\{~\}$ brackets.

\begin{marginfigure}[25mm]
\centering
  \begin{tikzpicture}
    \Vertices[style={shading=ball,blue}]{data/vertices.csv}
  \end{tikzpicture}
\end{marginfigure}

\begin{docspec}
\begin{lstlisting}
  \begin{tikzpicture}
    \Vertices[style={shading=ball,blue}]{data/vertices.csv}
  \end{tikzpicture}
\end{lstlisting}
\end{docspec}

\begin{docspecdef}
  \doccmd{Vertices}[\docopt{IdAsLabel}]\{\docarg{filename}\}
\end{docspecdef}
\begin{docspecdef}
  \doccmd{Vertices}[\docopt{NoLabel}]\{\docarg{filename}\}
\end{docspecdef}

\docopt{IdAsLabel} is a Boolean option which assigns the \docopt{id} of the single vertices as labels. On the contrary, \docopt{NoLabel} suppress all labels.

\begin{marginfigure}[22mm]
\centering
  \begin{tikzpicture}
    \Vertices[IdAsLabel]{data/vertices.csv}
    \node at (2,1)[font=\scriptsize]{C};
  \end{tikzpicture}
\end{marginfigure}

\begin{docspec}
\begin{lstlisting}
  \begin{tikzpicture}
    \Vertices[IdAsLabel]{data/vertices.csv}
  \end{tikzpicture}
\end{lstlisting}
\end{docspec}

\begin{marginfigure}[2mm]
\centering
  \begin{tikzpicture}
    \Vertices[NoLabel]{data/vertices.csv}
  \end{tikzpicture}
\end{marginfigure}

\begin{docspec}
\begin{lstlisting}
  \begin{tikzpicture}
    \Vertices[NoLabel]{data/vertices.csv}
  \end{tikzpicture}
\end{lstlisting}
\end{docspec}

\begin{docspecdef}
  \doccmd{Vertices}[\docopt{RGB}]\{\docarg{filename}\}
\end{docspecdef}

In order to display RGB colors for the vertex fill colors, the option \docopt{RGB} has to be entered. Additionally, the RGB values have to be specified in the file where the vertices are stored. Each value has its own column with the caption \docopt{R}, \docopt{G}, and \docopt{B}.

\marginnote[8mm]{File: vertices\_RGB.csv}
\begin{docspec}
\begin{lstlisting}
id, x, y ,size, color,opacity,label, R , G , B
 A, 0,  0, .4 , green,  .9   ,  a  ,255,  0,  0
 B, 1, .7, .6 ,      ,  .5   ,  b  ,  0,255,  0
 C, 2,  1, .8 ,orange,  .3   ,  c  ,  0,  0,255
 D, 2,  0, .5 ,   red,  .7   ,  d  , 10,120,255
 E,.2,1.5, .5 ,  gray,       ,  e  , 76, 55,255
\end{lstlisting}
\end{docspec}

The ``normal'' color definition can also be part of the vertex definition. If the option \docopt{RGB} is not set, then the colors under \docopt{color} are applied.

\newpage
\begin{marginfigure}[3mm]
\centering
  \begin{tikzpicture}
    \Vertices[RGB]{data/vertices_RGB.csv}
  \end{tikzpicture}
\end{marginfigure}

\begin{docspec}
\begin{lstlisting}
  \begin{tikzpicture}
    \Vertices[RGB]{data/vertices_RGB.csv}
  \end{tikzpicture}
\end{lstlisting}
\end{docspec}

\begin{docspecdef}
  \doccmd{Vertices}[\docopt{Pseudo}]\{\docarg{filename}\}
\end{docspecdef}

The option \docopt{Pseudo} creates a pseudo vertices, where only the names and the coordinates of the vertices wil be drawn. Edges etc, can still be assigned to these vertices.

\begin{docspecdef}
  \doccmd{Vertices}[\docopt{layer}=\docarg{number}]\{\docarg{filename}\}
\end{docspecdef}

With the option \docopt{layer}, only the vertices on the selected layer are plotted. More about this option and the use of layers is explained in Chapter~\ref{chap:multilayer_networks}.

\newpage
\section{Edges}
\label{sec:edges}

The \doccmddef{Edges} command is the extension of the \doccmd{Edge} command. Instead of a single edge, a set of edges will be drawn. This set of edges is defined in an external file but can be modified with \doccmd{Edges}.
\begin{docspecdef}
  \doccmd{Edges}[\docopt{global options}]\{\docarg{filename}\}
\end{docspecdef}

Like the vertices, the edges have to be stored in a clear text file\footnote{e.g. .txt, .tex, .csv, .dat, \dots}, preferentially in a \texttt{.csv} format. The first row should contain the headings, which are equal to the options defined in Table \ref{tab:edge_options}. Option are separated by a comma <<,>>. Each new row is corresponds to a new edge.

\marginnote[8mm]{File: edges.csv}
\begin{docspec}
\begin{lstlisting}
u,v,label,lw,color ,opacity,bend, R , G , B ,Direct
A,B, ab  ,.5,red   ,   1   ,  30,  0,120,255,false
B,C, bc  ,.7,blue  ,   1   , -60, 76, 55,255,false
B,D, bd  ,.5,blue  ,  .5   , -60, 76, 55,255,false
A,E, ae  , 1,green ,   1   ,  75,255,  0,  0,true
C,E, ce  , 2,orange,   1   ,   0,150,150,150,false
A,A, aa  ,.3,black ,  .5   ,  75,255,  0  ,0,false
\end{lstlisting}
\end{docspec}

The mandatory values are the \docopt{u} and \docopt{v} argument, which corresponds to the \docarg{Vertex i} and \docarg{Vertex j} arguments of a single \doccmd{Edge}. Edges can only create if a vertex exists with the same \docarg{Name}. All other options are optional. No specific order of the options must be maintained. If no value is entered for an option, the default value will be chosen\footnote{\TODO This is NOT valid for Boolean options, here values for all vertices have to be entered.}. The \docarg{filename} should not contain blank spaces or special characters. The edges are drawn by the command \doccmd{Edges} with the \docarg{filename} plus file format (e.g. \texttt{.csv}). If the edges file is not in the same directory as the main \LaTeX~file, also the path has to be specified. In order to draw edges, first, the vertices have to be generated. Only then, edges can be assigned.

\begin{marginfigure}[18mm]
\centering
  \begin{tikzpicture}
    \Vertices{data/vertices.csv}
    \Edges{data/edges.csv}
  \end{tikzpicture}
\end{marginfigure}

\begin{docspec}
\begin{lstlisting}
  \begin{tikzpicture}
    \Vertices{data/vertices.csv}
    \Edges{data/edges.csv}
  \end{tikzpicture}
\end{lstlisting}
\end{docspec}

Predefined \doccmd{Edge} options can be overruled by the \docopt{global options} of the \doccmd{Edges} command; I.e. these options apply for all edges in the file. For the \doccmd{Edges} the following options are available:

\newpage
\begin{table}[h]\index{Edges!options}
  \footnotesize%
  \begin{center}
    \begin{tabular}{lccl}
      \toprule
      Option & Default & Type &Definition \\
      \midrule
      lw          & \{\}  & measure & line width of the edge \\
      color       & \{\}  & color   & edge color\\
      opacity     & \{\}  & number  & opacity of the edge \\
      style       & \{\}  & string  & additional \tikzsym styles \\
      vertices    & \{\}  & file    & vertices were the edges are assigned to\\
      layer       & \{\}  & number  & edges in specific layers \\
      \midrule
      Direct      & false & Boolean & allow directed edges \\
      Math        & false & Boolean & displays the labels in math mode \\
      NoLabel     & false & Boolean & delete the labels\\
      RGB         & false & Boolean & allow RGB colors \\
      NotInBG     & false & Boolean & edges are not in the background layer\\
      \bottomrule
    \end{tabular}
    \scriptsize
  \end{center}
  \caption{Global options for the \doccmd{Edges} command.}
  \label{tab:edges_options}
\end{table}

The use of these options are similar to the options for a single \doccmd{Edge} defined in Section~\ref{sec:edge}.

\begin{docspecdef}
  \doccmd{Edges}[\docopt{lw}=\docarg{measure}]\{\docarg{filename}\}
\end{docspecdef}

The line width of the edges can be modified with the option \docopt{lw}. Here, the unit of the \docarg{measure} can be specified, otherwise, it is in \si{pt}.

\begin{marginfigure}[35mm]
\centering
  \begin{tikzpicture}
    \Vertices{data/vertices.csv}
    \Edges[lw=2.5]{data/edges.csv}
  \end{tikzpicture}
\end{marginfigure}

\begin{docspec}
\begin{lstlisting}
  \begin{tikzpicture}
    \Vertices{data/vertices.csv}
    \Edges[lw=2.5]{data/edges.csv}
  \end{tikzpicture}
\end{lstlisting}
\end{docspec}

\begin{docspecdef}
  \doccmd{Edges}[\docopt{color}=\docarg{color}]\{\docarg{filename}\}
\end{docspecdef}

To change the line color of all edges, the option \docopt{color} has to be used. Without the option \docopt{RGB} set, the default \tikzsym and \LaTeX~ colors can be applied.

\begin{marginfigure}[20mm]
\centering
  \begin{tikzpicture}
    \Vertices{data/vertices.csv}
    \Edges[color=green!70!blue]{data/edges.csv}
  \end{tikzpicture}
\end{marginfigure}

\begin{docspec}
\begin{lstlisting}
  \begin{tikzpicture}
    \Vertices{data/vertices.csv}
    \Edges[color=green!70!blue]{data/edges.csv}
  \end{tikzpicture}
\end{lstlisting}
\end{docspec}

\begin{docspecdef}
  \doccmd{Edges}[\docopt{opacity}=\docarg{number}]\{\docarg{filename}\}
\end{docspecdef}

With the option \docopt{opacity} the opacity of all edge lines can be modified. The range of the \docarg{number} lies between $0$ and $1$. Where $0$ represents a fully transparent fill and $1$ a solid fill.

\begin{marginfigure}[20mm]
\centering
  \begin{tikzpicture}
    \Vertices{data/vertices.csv}
    \Edges[opacity=0.3]{data/edges.csv}
  \end{tikzpicture}
\end{marginfigure}

\begin{docspec}
\begin{lstlisting}
  \begin{tikzpicture}
    \Vertices{data/vertices.csv}
    \Edges[opacity=0.3]{data/edges.csv}
  \end{tikzpicture}
\end{lstlisting}
\end{docspec}

\newpage
\begin{docspecdef}
  \doccmd{Edges}[\docopt{style}=\docarg{string}]\{\docarg{filename}\}
\end{docspecdef}

Any other \tikzsym style option or command can be entered via the option \docopt{style}. Most of these commands can be found in the ``\tikzsym and PGF Manual''.

\begin{marginfigure}[25mm]
\centering
  \begin{tikzpicture}
    \Vertices{data/vertices.csv}
    \Edges[style={dashed}]{data/edges.csv}
  \end{tikzpicture}
\end{marginfigure}

\begin{docspec}
\begin{lstlisting}
  \begin{tikzpicture}
    \Vertices{data/vertices.csv}
    \Edges[style={dashed}]{data/edges.csv}
  \end{tikzpicture}
\end{lstlisting}
\end{docspec}

\begin{docspecdef}
  \doccmd{Edges}[\docopt{Direct}]\{\docarg{filename}\}
\end{docspecdef}

Directed edges are created by enabling the option \docopt{Direct}. The arrow is drawn from \docopt{u} to \docopt{v}.

\begin{marginfigure}[15mm]
\centering
  \begin{tikzpicture}
    \Vertices{data/vertices.csv}
    \Edges[Direct]{data/edges.csv}
  \end{tikzpicture}
\end{marginfigure}

\begin{docspec}
\begin{lstlisting}
  \begin{tikzpicture}
    \Vertices{data/vertices.csv}
    \Edges[Direct]{data/edges.csv}
  \end{tikzpicture}
\end{lstlisting}
\end{docspec}

\begin{docspecdef}
  \doccmd{Edges}[Math]\{\docarg{filename}\}
\end{docspecdef}

The option \docopt{Math} allows transforming labels into mathematical expressions without using the \$~\$ environment.

\begin{docspecdef}
  \doccmd{Edges}[\docopt{NoLabel}]\{\docarg{filename}\}
\end{docspecdef}

The option \docopt{NoLabel} suppress all edge labels.

\begin{marginfigure}[30mm]
\centering
  \begin{tikzpicture}
    \Vertices{data/vertices.csv}
    \Edges[NoLabel]{data/edges.csv}
  \end{tikzpicture}
\end{marginfigure}

\begin{docspec}
\begin{lstlisting}
  \begin{tikzpicture}
    \Vertices{data/vertices.csv}
    \Edges[NoLabel]{data/edges.csv}
  \end{tikzpicture}
\end{lstlisting}
\end{docspec}

\begin{docspecdef}
  \doccmd{Edges}[\docopt{RGB}]\{\docarg{filename}\}
\end{docspecdef}

In order to display RGB colors for the edge line colors, the option \docopt{RGB} has to be entered. Additionally, the RGB values have to be specified in the file where the vertices are stored. Each value has its own column with the caption \docopt{R}, \docopt{G}, and \docopt{B}. The ``normal'' color definition can also be part of the vertex definition. If the option \docopt{RGB} is not set, then the colors under \docopt{color} are applied.

\begin{marginfigure}[34mm]
\centering
  \begin{tikzpicture}
    \Vertices{data/vertices.csv}
    \Edges[RGB]{data/edges.csv}
  \end{tikzpicture}
\end{marginfigure}

\begin{docspec}
\begin{lstlisting}
  \begin{tikzpicture}
    \Vertices{data/vertices.csv}
    \Edges[RGB]{data/edges.csv}
  \end{tikzpicture}
\end{lstlisting}
\end{docspec}

\begin{docspecdef}
  \doccmd{Edges}[\docopt{NotInBG}]\{\docarg{filename}\}
\end{docspecdef}

Per default, the edges are drawn on the background layer of the \docarg{tikzpicture}. I.e. objects which are created after the edges appear also on top of them. To turn this off, the option \docopt{NotInBG} has to be enabled.

\begin{docspecdef}
  \doccmd{Edges}[\docopt{vertices}=\docarg{filename}]\{\docarg{filename}\}
\end{docspecdef}

Edges can be assigned to a specific set of \doccmd{Vertices} with the option \docopt{vertices}. Thereby the argument \docarg{filename} is the same as used for the \doccmd{Vertices} command. This option might be necessary if multiple \doccmd{Vertices} are created and edges are assigned at the end.

\begin{docspecdef}
  \doccmd{Edges}[\docopt{layer}=\{\docarg{{layer $\alpha$}},\docarg{{layer $\beta$}}\}]\{\docarg{filename}\}
\end{docspecdef}

With the option \docopt{layer} only the edges between layer $\alpha$ and $\beta$ are plotted. The argument is a tuple of both layers indicated by  $\{~,~\}$. More about this option and the use of layers is explained in Chapter~\ref{chap:multilayer_networks}.

\chapter{Multilayer Networks}
\label{chap:multilayer_networks}

One of the main purposes of the \pkg package is the illustration of multilayer network structures. Thereby, all the previous commands can be used. A multilayer network is represented as a three-dimensional object, where each layer is located at a different $z$ plane. In order to enable this functionality, the option \docopt{multilayer} has to be used at the beginning of the \docenvdef{tikzpicture}.

\section{Simple Networks}
\label{sec:simple_networks}

\begin{docspecdef}
  \doccmd{Vertex}[\docopt{layer}=\docarg{number}]\{\docarg{Name}\}
\end{docspecdef}

With the option \docopt{layer} the vertex can be assigned to a specific layer. Layers are defined by numbers (e.g. $1$, $2$, $3$,\dots). Working with the \docopt{multilayer} option, each \doccmd{Vertex} has to be assigned to a specific layer. For the edge assignment no additional information is needed.

\begin{marginfigure}[45mm]
\centering
  \begin{tikzpicture}[multilayer]
    \begin{Layer}[layer=2]
      \draw[orange,very thick] (0,0) rectangle (2.5,1);
      \draw[step=.5, orange,draw opacity=.5] (0,0) grid (2.5,1);
    \end{Layer}
    \Vertex[x=.5,y=.5,IdAsLabel,layer=1]{A}
    \Vertex[x=1.5,y=.5,IdAsLabel,layer=1]{B}
    \Vertex[x=1.5,y=.5,IdAsLabel,layer=2]{C}
    \Edge[bend=60](A)(B)
    \Edge(C)(C)
    \Edge[style=dashed](B)(C)
  \end{tikzpicture}
\end{marginfigure}

\begin{docspec}
\begin{lstlisting}
  \begin{tikzpicture}[multilayer]
    \Vertex[x=0.5,IdAsLabel,layer=1]{A}
    \Vertex[x=1.5,IdAsLabel,layer=1]{B}
    \Vertex[x=1.5,IdAsLabel,layer=2]{C}
    \Edge[bend=60](A)(B)
    \Edge[style=dashed](B)(C)
    \Edge(C)(C)
  \end{tikzpicture}
\end{lstlisting}
\end{docspec}

Enabling the option \docopt{multilayer}, returns the network in a two-dimensional plane, like the networks discussed before. Setting the argument \docopt{multilayer}=\docarg{3d}, the network is rendered in a three-dimensional representation. Per default, the layer with the lowest number is on the top. This and the spacing between the layers can be changed with the command \doccmd{SetLayerDistance}.

\begin{marginfigure}[45mm]
\centering
  \begin{tikzpicture}[multilayer=3d]
    \SetLayerDistance{-1.5}
    \begin{Layer}[layer=1]
      \draw[orange,very thick] (0,0) rectangle (2.5,1);
      \draw[step=.5, orange,draw opacity=.5] (0,0) grid (2.5,1);
      \node at (0,0)[below right,orange]{Layer 1};
    \end{Layer}
    \begin{Layer}[layer=2]
      \draw[orange,very thick] (0,0) rectangle (2.5,1);
      \draw[step=.5, orange,draw opacity=.5] (0,0) grid (2.5,1);
      \node at (0,0)[below right,orange]{Layer 2};
    \end{Layer}
    \Vertex[x=0.5,y=.5,IdAsLabel,layer=1]{A}
    \Vertex[x=1.5,y=.5,IdAsLabel,layer=1]{B}
    \Vertex[x=1.5,y=.5,IdAsLabel,layer=2]{C}
    \Edge[bend=60](A)(B)
    \Edge(C)(C)
    \Edge[style=dashed](B)(C)
  \end{tikzpicture}
\end{marginfigure}

\begin{docspec}
\begin{lstlisting}
  \begin{tikzpicture}[multilayer=3d]
    \Vertex[x=0.5,IdAsLabel,layer=1]{A}
    \Vertex[x=1.5,IdAsLabel,layer=1]{B}
    \Vertex[x=1.5,IdAsLabel,layer=2]{C}
    \Edge[bend=60](A)(B)
    \Edge[style=dashed](B)(C)
    \Edge(C)(C)
  \end{tikzpicture}
\end{lstlisting}
\end{docspec}

\section{Complex Networks}

Similar as in Chapter \ref{chap:complex_networks} introduced, layers can be assigned to the vertices by adding a column \docopt{layer} to the file where the vertices are stored.

\marginnote[8mm]{File: ml\_vertices.csv}
\begin{docspec}
\begin{lstlisting}
id, x, y ,size, color,opacity,label,layer
 A, 0,  0, .4 , green,  .9   ,  a  ,  1
 B, 1, .7, .6 ,      ,  .5   ,  b  ,  1
 C, 2,  1, .8 ,orange,  .3   ,  c  ,  1
 D, 2,  0, .5 ,   red,  .7   ,  d  ,  2
 E,.2,1.5, .5 ,  gray,       ,  e  ,  1
 F,.1, .5, .7 ,  blue,  .3   ,  f  ,  2
 G, 2,  1, .4 ,  cyan,  .7   ,  g  ,  2
 H, 1,  1, .4 ,yellow,  .7   ,  h  ,  2
\end{lstlisting}
\end{docspec}

\marginnote[8mm]{File: ml\_edges.csv}
\begin{docspec}
\begin{lstlisting}
u,v,label,lw,color ,opacity,bend,Direct
A,B, ab  ,.5,red   ,   1   ,  30,false
B,C, bc  ,.7,blue  ,   1   , -60,false
A,E, ae  , 1,green ,   1   ,  45,true
C,E, ce  , 2,orange,   1   ,   0,false
A,A, aa  ,.3,black ,  .5   ,  75,false
C,G, cg  , 1,blue  ,  .5   ,   0,false
E,H, eh  , 1,gray  ,  .5   ,   0,false
F,A, fa  ,.7,red   ,  .7   ,   0,true
D,F, df  ,.7,cyan  ,   1   ,   30,true
F,H, fh  ,.7,purple,   1   ,   60,false
D,G, dg  ,.7,blue  ,  .7   ,   60,false
\end{lstlisting}
\end{docspec}

With the commands \doccmd{Vertices} and \doccmd{Edges}, the network can be created automatically. Again the \doccmd{Vertices} vertices should be performed first and then the command \doccmd{Edges}.

\begin{marginfigure}[25mm]
\centering
  \begin{tikzpicture}[multilayer=3d]
    \SetLayerDistance{-1.5}
    \Vertices[NoLabel,opacity=0]{data/ml_vertices.csv}
   \begin{Layer}[layer=2]
      \draw[orange,very thick] (-.5,-.5) rectangle (2.5,2);
      \draw[step=.5, orange,draw opacity=.5] (-.5,-.5) grid (2.5,2);
      \node at (-.5,-.5)[below right,orange]{Layer 2};
    \end{Layer}
    \Edges[NotInBG,layer={2,2}]{data/ml_edges.csv}
    \Edges[NotInBG,layer={1,2}]{data/ml_edges.csv}
    \Vertices[layer=2]{data/ml_vertices.csv}
    \begin{Layer}[layer=1]
      \draw[orange,very thick,fill=white,fill opacity=.7] (-.5,-.5) rectangle (2.5,2);
      \draw[step=.5, orange,draw opacity=.5] (-.5,-.5) grid (2.5,2);
      \node at (-.5,-.5)[below right,orange]{Layer 1};
    \end{Layer}
    \Edges[NotInBG,layer={1,1}]{data/ml_edges.csv}
    \Vertices[layer=1]{data/ml_vertices.csv}
  \end{tikzpicture}
\end{marginfigure}

\begin{docspec}
\begin{lstlisting}
  \begin{tikzpicture}[multilayer=3d]
    \Vertices{data/ml_vertices.csv}
    \Edges{data/ml_edges.csv}
  \end{tikzpicture}
\end{lstlisting}
\end{docspec}

\begin{docspecdef}
  \doccmd{Vertices}[\docopt{layer}=\docarg{number}]\{\docarg{filename}\}
\end{docspecdef}
\begin{docspecdef}
  \doccmd{Edges}[\docopt{layer}=\{\docarg{{layer $\alpha$}},\docarg{{layer $\beta$}}\}]\{\docarg{filename}\}
\end{docspecdef}

With the \doccmd{Vertices} option \docopt{layer} only the vertices on the selected layer are plotted. While, with the \doccmd{Edges} option \docopt{layer}, the edges between layer $\alpha$ and $\beta$ are plotted. The argument is a tuple of both layers indicated by  $\{~,~\}$.

\begin{marginfigure}[30mm]
\centering
  \begin{tikzpicture}[multilayer=3d]
    \SetLayerDistance{-1.5}
    \begin{Layer}[layer=1]
      \draw[orange,very thick,fill=white,fill opacity=.7] (-.5,-.5) rectangle (2.5,2);
      \draw[step=.5, orange,draw opacity=.5] (-.5,-.5) grid (2.5,2);
      \node at (-.5,-.5)[below right,orange]{Layer 1};
    \end{Layer}
    \Vertices[layer=1]{data/ml_vertices.csv}
    \Edges[NotInBG,layer={1,1}]{data/ml_edges.csv}
  \end{tikzpicture}
\end{marginfigure}

\begin{docspec}
\begin{lstlisting}
  \begin{tikzpicture}[multilayer=3d]
    \Vertices[layer=1]{data/ml_vertices.csv}
    \Edges[layer={1,1}]{data/ml_edges.csv}
  \end{tikzpicture}
\end{lstlisting}
\end{docspec}

\newpage

Plotting edges without defining first the vertices can be done with the \doccmd{Edges} option \docopt{vertices}. This allows modifying specific sets of Edges.

\begin{marginfigure}[15mm]
\centering
  \begin{tikzpicture}[multilayer=3d]
    \SetLayerDistance{-1.5}
    \Edges[vertices=data/ml_vertices.csv,layer={1,2},style=dashed]{data/ml_edges.csv}
  \end{tikzpicture}
\end{marginfigure}

\begin{docspec}
\begin{lstlisting}
  \begin{tikzpicture}[multilayer=3d]
    \Edges[vertices=data/ml_vertices.csv,
           layer={1,2},style=dashed]{data/ml_edges.csv}
  \end{tikzpicture}
\end{lstlisting}
\end{docspec}

\section{Layers and Layouts}

Besides adding vertices and edges to specific layers, every other \tikzsym object can be drawn on such a layer using the \docenvdef{Layer} environment. With the option \docopt{layer}=\docarg{layer $\alpha$}, the position of the canvas can be assigned to the specific layer.

\begin{docspecdef}
  \doccmd{begin}\{\docenv{Layer}\}[\docopt{layer}=\docarg{layer $\alpha$}]

  \doccmd{end}\{\docenv{Layer}\}
\end{docspecdef}

\begin{marginfigure}[40mm]
\centering
  \begin{tikzpicture}[multilayer=3d]
    \SetLayerDistance{-1.5}
    \begin{Layer}[layer=1]
      \draw[very thick] (-.5,-.5) rectangle (2.5,2);
      \node at (-.5,-.5)[below right]{Layer 1};
    \end{Layer}
    \Vertices[layer=1]{data/ml_vertices.csv}
    \Edges[layer={1,1}]{data/ml_edges.csv}
  \end{tikzpicture}
\end{marginfigure}

\begin{docspec}
\begin{lstlisting}
  \begin{tikzpicture}[multilayer=3d]
    \begin{Layer}[layer=1]
      \draw[very thick] (-.5,-.5) rectangle (2.5,2);
      \node at (-.5,-.5)[below right]{Layer 1};
    \end{Layer}
    \Vertices[layer=1]{data/ml_vertices.csv}
    \Edges[layer={1,1}]{data/ml_edges.csv}
  \end{tikzpicture}
\end{lstlisting}
\end{docspec}

\begin{docspecdef}
  \doccmd{SetLayerDistance}\{\docarg{measure}\}
\end{docspecdef}

With the command \doccmddef{SetLayerDistance} the distance between the layers and their orientation can be modified. Per default the distance is set to $-2$\doccmd{DefaultUnit} (here \si{cm}). A negative number implies that layers with a higher number will be stacked below layers with a smaller number.

\begin{docspecdef}
  \doccmd{SetCoordinates}[\docopt{xAngle}=\docarg{number},\docopt{yAngle}=\docarg{number},\docopt{zAngle}=\docarg{number}, \docopt{xLength}=\docarg{number},\docopt{yLength}=\docarg{number},\docopt{zLength}=\docarg{number}]
\end{docspecdef}

The perspective of the three-dimensional plot can be modified by changing the orientation of the coordinate system, which is done with the command \doccmddef{SetCoordinates}. Here the angle and the length of each axis can be modified. Angles are defined as a \docarg{number} in the range between $-360$ and $360$. Per default, the lengths of the axes are defined by the identity matrix, i.e. no distortion. If the length ratio is changed $x$, $y$, and/or $z$ values are distorted. The \doccmd{SetCoordinates} command has to be entered before the \docopt{multilayer} option is called!

\begin{marginfigure}[25mm]
\centering
\SetCoordinates[xAngle=-30,yLength=1.2,xLength=.8]
  \begin{tikzpicture}[multilayer=3d]
    \SetLayerDistance{-1.5}
    \begin{Layer}[layer=1]
      \draw[orange,very thick,fill=white,fill opacity=.7] (-.5,-.5) rectangle (2.5,2);
      \draw[step=.5, orange,draw opacity=.5] (-.5,-.5) grid (2.5,2);
      \node at (-.5,-.5)[below right,orange]{Layer 1};
    \end{Layer}
    \Vertices[layer=1]{data/ml_vertices.csv}
    \Edges[NotInBG,layer={1,1}]{data/ml_edges.csv}
  \end{tikzpicture}
\end{marginfigure}

\begin{docspec}
\begin{lstlisting}
  \SetCoordinates[xAngle=-30,yLength=1.2,xLength=.8]
  \begin{tikzpicture}[multilayer=3d]
    \Vertices[layer=1]{data/ml_vertices.csv}
    \Edges[layer={1,1}]{data/ml_edges.csv}
  \end{tikzpicture}
\end{lstlisting}
\end{docspec}
\newpage
\section{Plane}

To support the illustration of multilayer networks, the background of the layer can be simply visualized with the command \doccmddef{Plane}, which allow to draw boundaries, grids and include images to the layer.
\begin{docspecdef}
  \doccmd{Plane}[\docopt{options}]
\end{docspecdef}
No obligatory arguments are needed. For a \doccmd{Plane} the following options are available:

\begin{table}[h]\index{Plane!options}
  \footnotesize%
  \begin{center}
    \begin{tabular}{lccl}
      \toprule
      Option & Default & Type &Definition \\
      \midrule
      x          & 0     & measure& x-coordinate of the origin\\
      y          & 0     & measure& y-coordinate of the origin\\
      width      & \SI{5}{cm} & measure& width of the plane\\
      height     & \SI{5}{cm} & measure& height of the plane\\
      color      & vertexfill  & color  & fill color of the plane \\
      opacity    & 0.3   & number & opacity of the fill color \\
      grid       & \{\}  & measure& spacing of the grid\\
      image      & \{\}  & file   & path to the image file \\
      style      & \{\}  & string & additional \tikzsym styles \\
      layer      & 1  & number & layer where the plane is located\\
      \midrule
      RGB        & false & Boolean & allow RGB colors \\
      NoFill     & false & Boolean & disable fill color \\
      NoBorder   & false & Boolean & disable border line \\
      ImageAndFill&false & Boolean & allow image and fill color \\
      InBG       & false & Boolean & plane is in the background layer\\
      \bottomrule
    \end{tabular}
    \scriptsize
    \\$^a$ either measure or string
  \end{center}
  \caption{Options for the \doccmd{Plane} command.}
  \label{tab:plane_options}
\end{table}

\begin{docspecdef}
 \doccmd{Plane}[\docopt{x}=\docarg{measure},\docopt{y}=\docarg{measure},\docopt{width}=\docarg{measure},\docopt{height}=\docarg{measure}]
\end{docspecdef}

A \doccmd{Plane} is a rectangle with origin (\docopt{x},\docopt{y}), a given \docopt{width} and \docopt{height}. The origin is defined in the left lower corner and per default $(0,0)$. The plane is default \SI{5}{cm} (width) by \SI{5}{cm} (height). This default options can be changed with \doccmd{SetPlaneWidth} and \doccmd{SetPlaneHeight}\footnote{See Section \ref{sec:plane_style}.}

\begin{marginfigure}
\centering
  \begin{tikzpicture}[multilayer=3d]
    \Plane[x=-.5,y=-.5,width=3,height=2.5]
    \begin{Layer}[layer=1]
    \filldraw [orange] (-.5,-.5) circle (3pt);
    \node [orange,anchor=south west] at (-.5,-.5) {origin};
    \draw[|<->|,orange] (-.5,-.8) --++ (3,0) node [pos=.5,below]{width};
    \draw[|<->|,orange] (2.8,-.5) --++ (0,2.5) node [pos=.5,below,sloped]{height};
    \end{Layer}
  \end{tikzpicture}
\end{marginfigure}

\begin{docspec}
\begin{lstlisting}
  \begin{tikzpicture}[multilayer=3d]
    \Plane[x=-.5,y=-.5,width=3,height=2.5]
  \end{tikzpicture}
\end{lstlisting}
\end{docspec}

\begin{docspecdef}
  \doccmd{Plane}[\docopt{color}=\docarg{color}]
\end{docspecdef}

To change the fill color of each plane individually, the option \docopt{color} has to be used. Without the option \docopt{RGB} set, the default \tikzsym and \LaTeX~ colors can be applied. Per default the default vertex color is used.

\begin{marginfigure}[22mm]
\centering
  \begin{tikzpicture}[multilayer=3d]
    \Plane[x=-.5,y=-.5,width=3,height=2.5,color=green!70!blue]
  \end{tikzpicture}
\end{marginfigure}

\begin{docspec}
\begin{lstlisting}
  \begin{tikzpicture}[multilayer=3d]
    \Plane[x=-.5,y=-.5,width=3,height=2.5,color=green!70!blue]
  \end{tikzpicture}
\end{lstlisting}
\end{docspec}

\newpage

\begin{docspecdef}
  \doccmd{Plane}[\docopt{opacity}=\docarg{number}]
\end{docspecdef}

With the option \docopt{opacity} the opacity of the plane fill color can be modified. The range of the \docarg{number} lies between $0$ and $1$. Where $0$ represents a fully transparent fill and $1$ a solid fill. Per default the opacity is set to $0.3$.

\begin{marginfigure}[22mm]
\centering
  \begin{tikzpicture}[multilayer=3d]
    \Plane[x=-.5,y=-.5,width=3,height=2.5,opacity=.7]
  \end{tikzpicture}
\end{marginfigure}

\begin{docspec}
\begin{lstlisting}
  \begin{tikzpicture}[multilayer=3d]
    \Plane[x=-.5,y=-.5,width=3,height=2.5,opacity=.7]
  \end{tikzpicture}
\end{lstlisting}
\end{docspec}

\begin{docspecdef}
  \doccmd{Plane}[\docopt{grid}=\docarg{measure}]
\end{docspecdef}

With the option \docopt{grid} a grid will be drawn on top of the plane. The argument of this option defines the spacing between the grid lines. The entered \docarg{measures} are in default units (\si{cm}). Changing the unites (locally) can be done by adding the unit to the \docarg{measure}\footnote{e.g. x=\SI{5}{mm}}. Changes to the default setting can be made with \doccmd{SetDefaultUnit}\footnote{see Section \ref{sec:gerneral_settings}}.

\begin{marginfigure}
\centering
  \begin{tikzpicture}[multilayer=3d]
    \Plane[x=-.5,y=-.5,width=3,height=2.5,grid=5mm]
    \begin{Layer}[layer=1]
    \draw[<->,orange] (2,-.7) --++ (.5,0) node [pos=.5,below]{5mm};
    \draw[<->,orange] (2.7,-.5) --++ (0,.5) node [pos=.5,below,sloped]{5mm};
    \end{Layer}
  \end{tikzpicture}
\end{marginfigure}

\begin{docspec}
\begin{lstlisting}
  \begin{tikzpicture}[multilayer=3d]
    \Plane[x=-.5,y=-.5,width=3,height=2.5,grid=5mm]
  \end{tikzpicture}
\end{lstlisting}
\end{docspec}

\begin{docspecdef}
  \doccmd{Plane}[\docopt{image}=\docarg{file}]
\end{docspecdef}

An image can be assigned to a plane with the option \docopt{image}. The argument is the file name and the folder where the image is stored. The width and height of the figure is scaled to the size of the plane. Without the option \docopt{ImageAndFill} the image overwrite the color options.

\begin{marginfigure}[22mm]
\centering
  \begin{tikzpicture}[multilayer=3d]
    \Plane[x=-.5,y=-.5,width=3,height=2.5,image=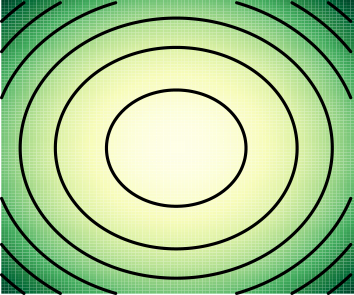]
  \end{tikzpicture}
\end{marginfigure}

\begin{docspec}
\begin{lstlisting}
  \begin{tikzpicture}[multilayer=3d]
    \Plane[x=-.5,y=-.5,width=3,height=2.5,image=data/plane.png]
  \end{tikzpicture}
\end{lstlisting}
\end{docspec}

\begin{docspecdef}
  \doccmd{Plane}[\docopt{style}=\docarg{string}]
\end{docspecdef}

Any other \tikzsym style option or command can be entered via the option \docopt{style}. Most of these commands can be found in the ``\tikzsym and PGF Manual''. Contain the commands additional options (e.g.\docopt{inner color}=\docarg{color}), then the argument for the \docopt{style} has to be between $\{~\}$ brackets.

\begin{marginfigure}[22mm]
\centering
  \begin{tikzpicture}[multilayer=3d]
    \Plane[x=-.5,y=-.5,width=3,height=2.5,style={dashed,inner color=white,outer color=red!80}]
  \end{tikzpicture}
\end{marginfigure}

\begin{docspec}
\begin{lstlisting}
  \begin{tikzpicture}[multilayer=3d]
    \Plane[x=-.5,y=-.5,width=3,height=2.5,style={dashed,inner color=white,outer color=red!80}]
  \end{tikzpicture}
\end{lstlisting}
\end{docspec}

\begin{docspecdef}
  \doccmd{Plane}[\docopt{layer}=\docarg{number}]
\end{docspecdef}

With the option \docopt{layer}=\docarg{layer $\alpha$}, the position of the plane can be assigned to a specific layer. Per default the plane is drawn on layer $1$.

\newpage
\begin{marginfigure}
\centering
  \begin{tikzpicture}[multilayer=3d]
    \SetLayerDistance{-1.5}
    \Plane[x=-.5,y=-.5,width=3,height=2.5,color=green,layer=2]
    \Plane[x=-.5,y=-.5,width=3,height=2.5]
    \Text[x=-.5,y=-.5,layer=2,color=orange,anchor=north west,fontsize=\normalsize]{Layer 2}
    \Text[x=-.5,y=-.5,layer=1,color=orange,anchor=north west,fontsize=\normalsize]{Layer 2}
  \end{tikzpicture}
\end{marginfigure}

\begin{docspec}
\begin{lstlisting}
  \begin{tikzpicture}[multilayer=3d]
    \SetLayerDistance{-1.5}
    \Plane[x=-.5,y=-.5,width=3,height=2.5,color=green,layer=2]
    \Plane[x=-.5,y=-.5,width=3,height=2.5]
  \end{tikzpicture}
\end{lstlisting}
\end{docspec}

\begin{docspecdef}
  \doccmd{Plane}[\docopt{RGB},\docopt{color}=\docarg{RGB values}]
\end{docspecdef}

In order to display RGB colors for the plane fill color, the option \docopt{RGB} has to be entered. In combination with this option, the \docopt{color} hast to be a list with the \docarg{RGB
values}, separated by <<,>> and within $\{~\}$.\footnote{e.g. the RGB code for white: $\{255,255,255\}$}

\begin{marginfigure}
\centering
  \begin{tikzpicture}[multilayer=3d]
    \Plane[x=-.5,y=-.5,width=3,height=2.5,RGB,color={0,0,0}]
  \end{tikzpicture}
\end{marginfigure}

\begin{docspec}
\begin{lstlisting}
  \begin{tikzpicture}[multilayer=3d]
    \Plane[x=-.5,y=-.5,width=3,height=2.5,RGB,color={0,0,0}]
  \end{tikzpicture}
\end{lstlisting}
\end{docspec}

\begin{docspecdef}
  \doccmd{Plane}[\docopt{NoFill}]
\end{docspecdef}
\begin{docspecdef}
  \doccmd{Plane}[\docopt{NoBorder}]
\end{docspecdef}

\docopt{NoFill} is a Boolean option which disables the fill color of the plane and \docopt{NoBorder} is a Boolean option which suppress the border line of the plane.

\begin{marginfigure}[16mm]
\centering
  \begin{tikzpicture}[multilayer=3d]
    \SetLayerDistance{-1.5}
    \Plane[x=-.5,y=-.5,width=3,height=2.5,layer=2,NoFill]
    \Plane[x=-.5,y=-.5,width=3,height=2.5,NoBorder]
    \Text[x=-.5,y=-.5,layer=2,color=orange,anchor=north west,fontsize=\normalsize]{Layer 2}
    \Text[x=-.5,y=-.5,layer=1,color=orange,anchor=north west,fontsize=\normalsize]{Layer 2}
  \end{tikzpicture}
\end{marginfigure}

\begin{docspec}
\begin{lstlisting}
  \begin{tikzpicture}[multilayer=3d]
    \SetLayerDistance{-1.5}
    \Plane[x=-.5,y=-.5,width=3,height=2.5,layer=2,NoFill]
    \Plane[x=-.5,y=-.5,width=3,height=2.5,NoBorder]
  \end{tikzpicture}
\end{lstlisting}
\end{docspec}

\begin{docspecdef}
  \doccmd{Plane}[\docopt{ImageAndFill}]
\end{docspecdef}

With the option \docopt{ImageAndFill} both, image and fill color can be drawn on a plane. The option \docopt{opacity} is applied to both objects.

\begin{marginfigure}[18mm]
\centering
  \begin{tikzpicture}[multilayer=3d]
    \Plane[x=-.5,y=-.5,width=3,height=2.5,image=data/plane.png,color=red,opacity=.4,ImageAndFill]
  \end{tikzpicture}
\end{marginfigure}

\begin{docspec}
\begin{lstlisting}
  \begin{tikzpicture}[multilayer=3d]
   \Plane[x=-.5,y=-.5,width=3,height=2.5,image=data/plane.png,color=red,opacity=.4,ImageAndFill]
  \end{tikzpicture}
\end{lstlisting}
\end{docspec}

\begin{docspecdef}
  \doccmd{Plane}[\docopt{InBG}]
\end{docspecdef}

A plane is drawn on the current layer of the \docarg{tikzpicture}. I.e. objects which are created after the plane appear on top of it and objects created before below of it. With the option \docopt{InBG} enabled, the plane is drawn on the background layer of the \docarg{tikzpicture}.

\chapter{Default Settings}
\label{chap:default_settings}

In order to customize the look of the networks, each layout setting used can be modified and adapted. There are three categories: General settings, vertex style, and edge style.

\section{General Settings}
\label{sec:gerneral_settings}

With the general settings mainly the sizes, distances and measures of the networks can be modified.

\begin{docspecdef}
  \doccmd{SetDefaultUnit}\{\docarg{unit}\}
\end{docspecdef}

The command \doccmddef{SetDefaultUnit} allows to change the units used for drawing the network\footnote{Except the line width, which are defined in \si{pt}.}, including diameters of the vertices, $x$ and $y$ coordinates or the distance between the layers. The default unit is \si{cm}.

\begin{docspecdef}
  \doccmd{SetDistanceScale}\{\docarg{number}\}
\end{docspecdef}

With the command \doccmddef{SetDistanceScale}, the distance between the vertices can be scaled. Per default \SI{1}{cm} entered corresponds to \SI{1}{cm} drawn, i.e. \doccmd{SetDistanceScale}\{1\}. Decreasing or increasing the scale changes the drawing distances between the vertices.

\begin{docspecdef}
  \doccmd{SetLayerDistance}\{\docarg{measure}\}
\end{docspecdef}

With the command \doccmddef{SetLayerDistance} the distance between the layers and their orientation can be modified. Per default, the distance is set to $-2$. A negative number implies that layers with a higher number will be stacked below layers with a smaller number.

\begin{docspecdef}
  \doccmd{SetCoordinates}[\docopt{xAngle}=\docarg{number},\docopt{yAngle}=\docarg{number},\docopt{zAngle}=\docarg{number}, \docopt{xLength}=\docarg{number},\docopt{yLength}=\docarg{number},\docopt{zLength}=\docarg{number}]
\end{docspecdef}

The perspective of the three-dimensional plot can be modified by changing the orientation of the coordinate system, which is done with the command \doccmddef{SetCoordinates}. Here the angle and the length of each axis can be modified. Angles are defined as a \docarg{number} in the range between $-360$ and $360$. Per default, the length of the axes are defined by the identity matrix, i.e. no distortion. If the length ratio is changed $x$, $y$, and/or $z$ values are distorted. The \doccmd{SetCoordinates} command has to be entered before the \docopt{multilayer} option is called!

\section{Vertex Style}
\label{sec:vertex_style}

The appearance of the vertices can be modified with the command \doccmddef{SetVertexStyle}. This command will change the default settings of the vertices in the network.

\begin{docspecdef}
  \doccmd{SetVertexStyle}[\docarg{document options}]
\end{docspecdef}

The following options are available:

\begin{table}[h]\index{Vertex!style}
  \footnotesize%
  \begin{center}
    \begin{tabularx}{\textwidth}{lccX}
      \toprule
      Option & Default & Type &Definition \\
      \midrule
Shape        & circle                  & text     & shape of the vertex \\
InnerSep     & 2pt                     & measure  & separation space which will be added inside the shape \\
OuterSep     & 0pt                     & measure  & separation space outside the background path \\
MinSize      & 0.6\doccmd{DefaultUnit} & measure  & diameter (size) of the vertex \\
FillColor    & vertexfill              & color    & color of the vertex \\
FillOpacity  & 1                       & number   & opacity of the vertex \\
LineWidth    & 1pt                     & measure  & line width of the vertex boundary \\
LineColor    & black                   & color    & line color of the vertex boundary \\
LineOpacity  & 1                       & number   & line opacity of the vertex boundary \\
TextFont     & \doccmd{scriptsize}     & fontsize & font size of the vertex label \\
TextColor    & black                   & color    & color of the vertex label \\
TextOpacity  & 1                       & number   & opacity of the vertex label \\
TextRotation & 0                       & number   & initial rotation of the vertex \\
      \bottomrule
    \end{tabularx}
    \scriptsize
  \end{center}
  \caption{Document style options for the vertices.}
  \label{tab:vertex_style}
\end{table}

\section{Edge Style}
\label{sec:edge_style}

The appearance of the edges can be modified with the command \doccmddef{SetEdgeStyle}. This command will change the default settings of the edges in the network.

\begin{docspecdef}
  \doccmd{SetEdgeStyle}[\docarg{document options}]
\end{docspecdef}

The following options are available:

\begin{table}[h]\index{Edge!style}
  \footnotesize%
  \begin{center}
    \begin{tabularx}{\textwidth}{lccX}
      \toprule
      Option & Default & Type &Definition \\
      \midrule
LineWidth       & 1.5pt               & measure  & width of the edge \\
Color           & black!75            & color    & color of the edge \\
Opacity         & 1                   & number   & opacity of the edge \\
Arrow           & -latex              & text     & arrow shape of the directed edge \\
TextFont        & \doccmd{scriptsize} & fontsize & font size of the edge label \\
TextOpacity     & 1                   & number   & opacity of the edge label \\
TextFillColor   & white               & color    & fill color of the edge label \\
TextFillOpacity & 1                   & number   & fill opacity of the edge label \\
InnerSep        & 0pt                 & measure  & separation space which will be added inside the shape \\
OuterSep        & 1pt                 & measure  & separation space outside the background path \\
TextRotation    & 0                   & number   & initial rotation of the edge label \\
      \bottomrule
    \end{tabularx}
    \scriptsize
  \end{center}
  \caption{Document style options for the edges.}
  \label{tab:edge_style}
\end{table}

\begin{docspecdef}
  \doccmd{EdgesNotInBG}
\end{docspecdef}

\begin{docspecdef}
  \doccmd{EdgesInBG}
\end{docspecdef}

Per default edges are drawn on the background layer, with the command \doccmddef{EdgesNotInBG} this can be disabled, while the command \doccmddef{EdgesInBG} restores the default setting.

\section{Text Style}
\label{sec:text_style}

The appearance of the text can be modified with the command \doccmddef{SetTextStyle}. This command will change the default settings of the text.

\begin{docspecdef}
  \doccmd{SetTextStyle}[\docarg{document options}]
\end{docspecdef}

The following options are available:

\begin{table}[h]\index{Text!style}
  \footnotesize%
  \begin{center}
    \begin{tabularx}{\textwidth}{lccX}
      \toprule
      Option & Default & Type &Definition \\
      \midrule
TextFont        & \doccmd{normalsize} & fontsize & font size of the text \\
TextOpacity     & 1                   & number   & opacity of the text \\
TextColor       & black               & color    & color of the text \\
TextOpacity     & 1                   & number   & opacity of the text \\
InnerSep        & 2pt                 & measure  & separation space which will be added inside the shape \\
OuterSep        & 0pt                 & measure  & separation space outside the background path \\
TextRotation    & 0                   & number   & initial rotation of the text \\
      \bottomrule
    \end{tabularx}
    \scriptsize
  \end{center}
  \caption{Document style options for the planes.}
  \label{tab:plane_style}
\end{table}

\section{Plane Style}
\label{sec:plane_style}

The appearance of the planes can be modified with the command \doccmddef{SetPlaneStyle}. This command will change the default settings of the planes.

\begin{docspecdef}
  \doccmd{SetPlaneStyle}[\docarg{document options}]
\end{docspecdef}

The following options are available:

\begin{table}[h]\index{Plane!style}
  \footnotesize%
  \begin{center}
    \begin{tabularx}{\textwidth}{lccX}
      \toprule
      Option & Default & Type &Definition \\
      \midrule
LineWidth     & 1.5pt               & measure  & width of the border line \\
LineColor     & black               & color    & color of the border line \\
LineOpacity   & 1                   & number   & opacity of the border line \\
FillColor     & vertexfill          & color    & fill color of the plane \\
FillOpacity   & 0.3                 & number   & fill opacity of the plane \\
GridLineWidth & 0.5pt               & measure  & width of the grid lines \\
GridColor     & black               & color    & color of the grid lines \\
GridOpacity   & 0.5                 & number   & opacity of the grid lines \\
      \bottomrule
    \end{tabularx}
    \scriptsize
  \end{center}
  \caption{Document style options for the planes.}
  \label{tab:plane_style}
\end{table}

\begin{docspecdef}
  \doccmd{SetPlaneWidth}\{\docarg{measure}\}
\end{docspecdef}
\begin{docspecdef}
  \doccmd{SetPlaneHeight}\{\docarg{measure}\}
\end{docspecdef}

With the commands \doccmd{SetPlaneWidth} and \doccmd{SetPlaneHeight} the default size of the planes can be modified.

\chapter{Troubleshooting and Support}
\label{chap:troubleshooting}

\section{\pkg Website}\label{sec:website}
The website for the \pkg packages is located at
\url{https://github.com/hackl/tikz-network}.  There, you'll find the actual version of the source code, a bug tracker, and the documentation.

\section{Getting Help}\label{sec:getting-help}
If you've encountered a problem with one of the \pkg commands, have a
question, or would like to report a bug, please send an email to me or visit our website.

To help me troubleshoot the problem more quickly, please try to compile your
document using the \docclsopt{debug} class option and send the generated
\texttt{.log} file to the mailing list with a brief description of the problem.

\section{Errors, Warnings, and Informational Messages}\label{sec:tl-messages}
The following is a list of all of the errors, warnings, and other messages generated by the \pkg classes and a brief description of their meanings.
\index{error messages}\index{warning messages}\index{debug messages}

\docmsg{Error: ! TeX capacity exceeded, sorry [main memory size=5000000].}{%
The considered network is to large and \texttt{pdflatex} runs out of memory. This problem can be solved by using \texttt{lualatex} or \texttt{xetex} instead.}

\section{Package Dependencies}\label{sec:dependencies}
The following is a list of packages that the \pkg package rely upon.  Packages marked with an asterisk are optional.
\begin{multicols}{2}
  \begin{itemize}
  \item etex
  \item xifthen
  \item xkeyval
  \item datatool
  \item tikz
    \begin{itemize}
    \item arrows
    \item positioning
    \item 3d
    \item fit
    \item calc
    \item backgrounds
    \end{itemize}
  \end{itemize}
\end{multicols}

\appendix
\chapter{ToDo}
\label{chap:todo}

\section{Code to fix}
\begin{itemize}
\item change default entries for Boolean options in the vertices file.
\end{itemize}

\section{Documentation}
\begin{itemize}
\item add indices to the manual.
\item extended tutorial/example to the document.
\item clean-up and document the .sty file.
\end{itemize}

\section{Features}
\begin{itemize}
\item add a spherical coordinate system
\end{itemize}

\section{Add-ons}
\begin{itemize}
\item add QGIS to tikz-network compiler
\end{itemize}

\chapter{Add-ons}
\label{chap:todo}

\section{Python networks to \tikzsym with network2tikz}
\label{sec:python_to_tikz}

\subsection{Introduction}

\texttt{\href{https://github.com/hackl/network2tikz}{network2tikz}} is a Python tool for converting network visualizations into \pkg figures, for native inclusion into your LaTeX documents.

\texttt{network2tikz} works with Python 3 and supports (currently) the following Python network modules:

\begin{itemize}
\item \texttt{\href{https://github.com/hackl/cnet}{cnet}}
\item \texttt{\href{http://igraph.org/python/}{python-igraph}}
\item \texttt{\href{https://networkx.github.io/}{networkx}}
\item \texttt{\href{https://github.com/IngoScholtes/pathpy}{pathpy}}
\item default node/edge lists
\end{itemize}

The output of \texttt{network2tikz} is a \pkg figure. Because you are not only getting an image of your network, but also the LaTeX source file, you can easily post-process the figures (e.g. adding drawings, texts, equations,...).

Since \textit{a picture is worth a thousand words} a small example:

\begin{marginfigure}[65mm]
\centering
  \begin{tikzpicture}[scale=.8]
    \Vertex[x=0.785,y=2.375,color=red,opacity=0.5,label=Alice]{a}
    \Vertex[x=5.215,y=5.650,color=blue,opacity=0.5,label=Bob]{b}
    \Vertex[x=3.819,y=0.350,color=red,opacity=0.5,label=Claire]{c}
    \Vertex[x=4.654,y=2.051,color=blue,opacity=0.5,label=Dennis]{d}
    \Edge[,bend=-8.531](a)(c)
    \Edge[,bend=-8.531](c)(d)
    \Edge[,bend=-8.531](d)(b)
    \Edge[,bend=-8.531](a)(b)
  \end{tikzpicture}
\end{marginfigure}

\begin{docspec}
\begin{lstlisting}
#!/usr/bin/python -tt
# -*- coding: utf-8 -*-

nodes = ['a','b','c','d']
edges = [('a','b'), ('a','c'), ('c','d'),('d','b')]
gender = ['f', 'm', 'f', 'm']
colors = {'m': 'blue', 'f': 'red'}

style = {}
style['node_label'] = ['Alice', 'Bob', 'Claire', 'Dennis']
style['node_color'] = [colors[g] for g in gender]
style['node_opacity'] = .5
style['edge_curved'] = .1

from network2tikz import plot
plot((nodes,edges),'network.tex',**style)
\end{lstlisting}
\end{docspec}

(see above) gives

\begin{docspec}
\begin{lstlisting}
\documentclass{standalone}
\usepackage{tikz-network}
\begin{document}
\begin{tikzpicture}
\clip (0,0) rectangle (6,6);
\Vertex[x=0.785,y=2.375,color=red,opacity=0.5,label=Alice]{a}
\Vertex[x=5.215,y=5.650,color=blue,opacity=0.5,label=Bob]{b}
\Vertex[x=3.819,y=0.350,color=red,opacity=0.5,label=Claire]{c}
\Vertex[x=4.654,y=2.051,color=blue,opacity=0.5,label=Dennis]{d}
\Edge[,bend=-8.531](a)(c)
\Edge[,bend=-8.531](c)(d)
\Edge[,bend=-8.531](d)(b)
\Edge[,bend=-8.531](a)(b)
\end{tikzpicture}
\end{document}
\end{lstlisting}
\end{docspec}

Tweaking the plot is straightforward and can be done as part of your LaTeX workflow.

\subsection{Installation}

\texttt{network2tikz} is available from the \href{https://pypi.org/project/network2tikz/}{Python Package Index}, so simply type

\begin{docspec}
\begin{lstlisting}
pip install -U network2tikz
\end{lstlisting}
\end{docspec}

to install/update. If your are intersted in the development version of the module check out the \href{https://github.com/hackl/network2tikz}{github repository}.

\subsection{Usage}

\begin{enumerate}
\item Generate, manipulation, and study of the structure, dynamics, and functions of your complex networks as usual, with your preferred python module.

\item Instead of the default plot functions (e.g. \texttt{igraph.plot()} or \texttt{networkx.draw()}) invoke \texttt{network2tikz} by
\begin{docspec}
\begin{lstlisting}
plot(G,'mytikz.tex')
\end{lstlisting}
\end{docspec}
to store your network visualisation as the TikZ file \texttt{mytikz.tex}. Load the module with:
\begin{docspec}
\begin{lstlisting}
from network2tikz import plot
\end{lstlisting}
\end{docspec}

\textbf{Advanced usage:}

Of course, you always can improve your plot by manipulating the generated LaTeX file, but why not do it directly in Python? To do so, all visualization options available in \pkg are also implemented in \texttt{network2tikz}. The appearance of the plot can be modified by keyword arguments.\footnote{For a detailed explanation, please see Section \ref{sec:plot_function}.}
\begin{docspec}
\begin{lstlisting}
my_style = {}
plot(G,'mytikz.tex',**my_style)
\end{lstlisting}
\end{docspec}

The arguments follow the options described above in the manual.

Additionally, if you are more interested in the final output and  not only the \texttt{.tex} file, used
\begin{docspec}
\begin{lstlisting}
plot(G,'mypdf.pdf')
\end{lstlisting}
\end{docspec}
to save your plot as a pdf, or
\begin{docspec}
\begin{lstlisting}
plot(G)
\end{lstlisting}
\end{docspec}
to create a temporal plot and directly show the result, i.e. similar to the matplotlib function \texttt{show()}. Finally, you can also create a node and edge list, which can be read and easily modified (in a post-processing step) as showd above.
\begin{docspec}
\begin{lstlisting}
plot(G,'mycsv.csv')
\end{lstlisting}
\end{docspec}

\item Compile the figure or add the contents of \texttt{mytikz.tex} into your LaTeX source code. With the option \docopt{standalone}=\docarg{false} only the \tikzsym figure will be saved, which can then be easily included in your \LaTeX~ document via \doccmd{input}\{\texttt{/path/to/mytikz.tex}\}.

\subsection{Simple example}

For illustration purpose, a similar network as in the \href{http://igraph.org/python/doc/tutorial/tutorial.html}{python-igraph tutorial} is used. If you are using another Python network module, and like to follow this example, please have a look at the \href{https://github.com/hackl/network2tikz/tree/master/examples}{provided examples}.

\end{enumerate}

Create network object and add some edges.
\begin{docspec}
\begin{lstlisting}
#!/usr/bin/python -tt
# -*- coding: utf-8 -*-

import igraph
from network2tikz import plot

net = igraph.Graph([(0,1), (0,2), (2,3), (3,4), (4,2), (2,5), (5,0), (6,3),
                    (5,6), (6,6)],directed=True)
\end{lstlisting}
\end{docspec}

Adding node and edge properties.

\begin{docspec}
\begin{lstlisting}
net.vs["name"] = ["Alice", "Bob", "Claire", "Dennis", "Esther", "Frank", "George"]
net.vs["age"] = [25, 31, 18, 47, 22, 23, 50]
net.vs["gender"] = ["f", "m", "f", "m", "f", "m", "m"]
net.es["is_formal"] = [False, False, True, True, True, False, True, False,
                       False, False]
\end{lstlisting}
\end{docspec}

Already now the network can be plotted.
\begin{marginfigure}[5mm]
\centering
  \begin{tikzpicture}[scale=.8]
\Vertex[x=3.629,y=0.350]{a}
\Vertex[x=3.449,y=5.457]{b}
\Vertex[x=2.647,y=5.650]{c}
\Vertex[x=1.013,y=3.477]{d}
\Vertex[x=4.987,y=2.834]{e}
\Vertex[x=4.490,y=5.182]{f}
\Vertex[x=2.808,y=4.470]{g}
\Edge[](a)(b)
\Edge[](a)(c)
\Edge[](c)(d)
\Edge[](d)(e)
\Edge[](e)(c)
\Edge[](c)(f)
\Edge[](f)(a)
\Edge[](f)(g)
\Edge[](g)(d)
\Edge[](g)(g)
  \end{tikzpicture}
\end{marginfigure}

\begin{docspec}
\begin{lstlisting}
plot(net)
\end{lstlisting}
\end{docspec}

Per default, the node positions are assigned uniform random. In order to create a layout, the layout methods of the network packages can be used. Or the position of the nodes can be directly assigned, in form of a dictionary, where the key is the node id and the value is a tuple of the node position in $x$ and $y$.

\begin{docspec}
\begin{lstlisting}
layout = {0: (4.3191, -3.5352), 1: (0.5292, -0.5292),
          2: (8.6559, -3.8008), 3: (12.4117, -7.5239),
          4: (12.7, -1.7069), 5: (6.0022, -9.0323),
          6: (9.7608, -12.7)}
plot(net,layout=layout)
\end{lstlisting}
\end{docspec}

\begin{marginfigure}[5mm]
\centering
  \begin{tikzpicture}[scale=.8]
\clip (0,0) rectangle (6,6);
\Vertex[x=2.000,y=4.341]{a}
\Vertex[x=0.350,y=5.650]{b}
\Vertex[x=3.889,y=4.225]{c}
\Vertex[x=5.524,y=2.604]{d}
\Vertex[x=5.650,y=5.137]{e}
\Vertex[x=2.733,y=1.947]{f}
\Vertex[x=4.370,y=0.350]{g}
\Edge[](a)(b)
\Edge[](a)(c)
\Edge[](c)(d)
\Edge[](d)(e)
\Edge[](e)(c)
\Edge[](c)(f)
\Edge[](f)(a)
\Edge[](f)(g)
\Edge[](g)(d)
\Edge[](g)(g)
  \end{tikzpicture}
\end{marginfigure}

This should open an external pdf viewer showing a visual representation of the network, something like the one on the following figure:

We can simply re-using the previous layout object here, but we also specified that we need a bigger plot ($8 \times 8$ cm) and a larger margin around the graph to fit the self loop and potential labels (1 cm).\footnote{Per default, all size values are based on cm, and all line widths are defined in pt units. With the general option \docopt{units} this can be changed, see Section \ref{sec:plot_function}.}

\begin{docspec}
\begin{lstlisting}
plot(net, layout=layout, canvas=(8,8), margin=1)
\end{lstlisting}
\end{docspec}

\begin{marginfigure}[5mm]
\centering
  \begin{tikzpicture}[scale=.8]
\Vertex[x=2.868,y=5.518]{a}
\Vertex[x=1.000,y=7.000]{b}
\Vertex[x=5.006,y=5.387]{c}
\Vertex[x=6.858,y=3.552]{d}
\Vertex[x=7.000,y=6.419]{e}
\Vertex[x=3.698,y=2.808]{f}
\Vertex[x=5.551,y=1.000]{g}
\Edge[](a)(b)
\Edge[](a)(c)
\Edge[](c)(d)
\Edge[](d)(e)
\Edge[](e)(c)
\Edge[](c)(f)
\Edge[](f)(a)
\Edge[](f)(g)
\Edge[](g)(d)
\Edge[](g)(g)
  \end{tikzpicture}
\end{marginfigure}

In to keep the properties of the visual representation of your network separate from the network itself. You can simply set up a Python dictionary containing the keyword arguments you would pass to \texttt{plot} and then use the double asterisk (\texttt{**}) operator to pass your specific styling attributes to \texttt{plot}:

\begin{docspec}
\begin{lstlisting}
color_dict = {'m': 'blue', 'f': 'red'}
visual_style = {}

# Node options
visual_style['vertex_size'] = .5
visual_style['vertex_color'] = [color_dict[g] for g in net.vs['gender']]
visual_style['vertex_opacity'] = .7
visual_style['vertex_label'] = net.vs['name']
visual_style['vertex_label_position'] = 'below'

# Edge options
visual_style['edge_width'] = [1 + 2 * int(f) for f in net.es('is_formal')]
visual_style['edge_curved'] = 0.1

# General options and plot command.
visual_style['layout'] = layout
visual_style['canvas'] = (8,8)
visual_style['margin'] = 1

# Plot command
plot(net,**visual_style)
\end{lstlisting}
\end{docspec}

\begin{marginfigure}[5mm]
\centering
  \begin{tikzpicture}[scale=.8]
\Vertex[x=2.868,y=5.518,size=0.5,color=red,opacity=0.7,label=Alice,position=below]{a}
\Vertex[x=1.000,y=7.000,size=0.5,color=blue,opacity=0.7,label=Bob,position=below]{b}
\Vertex[x=5.006,y=5.387,size=0.5,color=red,opacity=0.7,label=Claire,position=below]{c}
\Vertex[x=6.858,y=3.552,size=0.5,color=blue,opacity=0.7,label=Dennis,position=below]{d}
\Vertex[x=7.000,y=6.419,size=0.5,color=red,opacity=0.7,label=Esther,position=below]{e}
\Vertex[x=3.698,y=2.808,size=0.5,color=blue,opacity=0.7,label=Frank,position=below]{f}
\Vertex[x=5.551,y=1.000,size=0.5,color=blue,opacity=0.7,label=George,position=below]{g}
\Edge[,lw=1.0,bend=-8.531,Direct](a)(b)
\Edge[,lw=1.0,bend=-8.531,Direct](a)(c)
\Edge[,lw=3.0,bend=-8.531,Direct](c)(d)
\Edge[,lw=3.0,bend=-8.531,Direct](d)(e)
\Edge[,lw=3.0,bend=-8.531,Direct](e)(c)
\Edge[,lw=1.0,bend=-8.531,Direct](c)(f)
\Edge[,lw=3.0,bend=-8.531,Direct](f)(a)
\Edge[,lw=1.0,bend=-8.531,Direct](f)(g)
\Edge[,lw=1.0,bend=-8.531,Direct](g)(g)
\Edge[,lw=1.0,bend=-8.531,Direct](g)(d)
  \end{tikzpicture}
\end{marginfigure}

Beside showing the network, we can also generate the latex source file, which can be used and modified later on. This is done by adding the output file name with the ending \texttt{'.tex'}.

\begin{docspec}
\begin{lstlisting}
plot(net,'network.tex',**visual_style)
\end{lstlisting}
\end{docspec}

produces

\begin{fullwidth}
\begin{minipage}{17cm}
\begin{docspec}
\begin{lstlisting}
\documentclass{standalone}
\usepackage{tikz-network}
\begin{document}
\begin{tikzpicture}
\clip (0,0) rectangle (8.0,8.0);
\Vertex[x=2.868,y=5.518,size=0.5,color=red,opacity=0.7,label=Alice,position=below]{a}
\Vertex[x=1.000,y=7.000,size=0.5,color=blue,opacity=0.7,label=Bob,position=below]{b}
\Vertex[x=5.006,y=5.387,size=0.5,color=red,opacity=0.7,label=Claire,position=below]{c}
\Vertex[x=6.858,y=3.552,size=0.5,color=blue,opacity=0.7,label=Dennis,position=below]{d}
\Vertex[x=7.000,y=6.419,size=0.5,color=red,opacity=0.7,label=Esther,position=below]{e}
\Vertex[x=3.698,y=2.808,size=0.5,color=blue,opacity=0.7,label=Frank,position=below]{f}
\Vertex[x=5.551,y=1.000,size=0.5,color=blue,opacity=0.7,label=George,position=below]{g}
\Edge[,lw=1.0,bend=-8.531,Direct](a)(b)
\Edge[,lw=1.0,bend=-8.531,Direct](a)(c)
\Edge[,lw=3.0,bend=-8.531,Direct](c)(d)
\Edge[,lw=3.0,bend=-8.531,Direct](d)(e)
\Edge[,lw=3.0,bend=-8.531,Direct](e)(c)
\Edge[,lw=1.0,bend=-8.531,Direct](c)(f)
\Edge[,lw=3.0,bend=-8.531,Direct](f)(a)
\Edge[,lw=1.0,bend=-8.531,Direct](f)(g)
\Edge[,lw=1.0,bend=-8.531,Direct](g)(g)
\Edge[,lw=1.0,bend=-8.531,Direct](g)(d)
\end{tikzpicture}
\end{document}
\end{lstlisting}
\end{docspec}
\end{minipage}
\end{fullwidth}

Instead of the tex file, a node and edge list can be generates, which can also be used with the library.

\begin{docspec}
\begin{lstlisting}
plot(net,'network.csv',**visual_style)
\end{lstlisting}
\end{docspec}

The node list \texttt{network\_nodes.csv}.
\begin{docspec}
\begin{lstlisting}
id,x,y,size,color,opacity,label,position
a,2.868,5.518,0.5,red,0.7,Alice,below
b,1.000,7.000,0.5,blue,0.7,Bob,below
c,5.006,5.387,0.5,red,0.7,Claire,below
d,6.858,3.552,0.5,blue,0.7,Dennis,below
e,7.000,6.419,0.5,red,0.7,Esther,below
f,3.698,2.808,0.5,blue,0.7,Frank,below
g,5.551,1.000,0.5,blue,0.7,George,below
\end{lstlisting}
\end{docspec}

The edge list \texttt{network\_edges.csv}.
\begin{docspec}
\begin{lstlisting}
u,v,lw,bend,Direct
a,b,1.0,-8.531,true
a,c,1.0,-8.531,true
c,d,3.0,-8.531,true
d,e,3.0,-8.531,true
e,c,3.0,-8.531,true
c,f,1.0,-8.531,true
f,a,3.0,-8.531,true
f,g,1.0,-8.531,true
g,g,1.0,-8.531,true
g,d,1.0,-8.531,true
\end{lstlisting}
\end{docspec}

\subsection{The plot function in detail}
\label{sec:plot_function}

\begin{docspecdef}
  \texttt{network2tikz.plot}(\docarg{network}, \docopt{filename}=\docarg{None}, \docopt{type}= \docarg{None}, \docopt{**kwds})
\end{docspecdef}

\textbf{Parameters}

\begin{description}
\item[network] : network object

Network to be drawn. The network can be a \texttt{cnet}, \texttt{networkx}, \texttt{igraph}, \texttt{pathpy} object, or a tuple of a node list and edge list.

\item[filename] : file, string or None, optional (default = None)

  File or filename to save. The file ending specifies the output. i.e. is the file ending with \texttt{.tex} a tex file will be created; if the file ends with '.pdf' a pdf is created; if the file ends with \texttt{.csv}, two csv files are generated \texttt{filename\_nodes.csv} and \texttt{filename\_edges.csv}. If the filename is a tuple of strings, the first entry will be used to name the node list and the second entry for the edge list; and if no ending and no type is defined a temporary pdf file is compiled and shown.

\item[type] : str or None, optional (default = None)

  Type of the output file. If no ending is defined trough the filename, the type of the output file can be specified by the type option. Currently the following output types are supported: \texttt{'tex'}, \texttt{'pdf'}, \texttt{'csv'} and \texttt{'dat'}.

\item[kwds] : keyword arguments, optional (default= no attributes)

    Attributes used to modify the appearance of the plot. For details see below.

\end{description}

\textbf{Keyword arguments for node styles}

\begin{description}
\item[node\_size] : size of the node. The default is 0.6 cm.

\item[node\_color] : color of the nodes. The default is light blue. Colors can
  be specified either by common color names, or by 3-tuples of floats
  (ranging between 0 and 255 for the R, G and B components).

\item[node\_opacity] : opacity of the nodes. The default is 1. The range of the
  number lies between 0 and 1. Where 0 represents a fully transparent fill
  and 1 a solid fill.

\item[node\_label] : labels drawn next to the nodes.

\item[node\_label\_position] : Per default the position of the label is in the
  center of the node. Classical \tikzsym commands can be used to change the
  position of the label. Instead, using such command, the position can be
  determined via an angle, by entering a number between -360 and 360. The
  origin (0) is the $y$ axis. A positive number change the position counter
  clockwise, while a negative number make changes clockwise.

\item[node\_label\_distance] : distance between the node and the label.

\item[node\_label\_color] : color of the label.

\item[node\_label\_size] : font size of the label.

\item[node\_shape] : shape of the vertices. Possibilities are:
  'circle', 'rectangle',  'triangle', and any other Tikz shape

\item[node\_style] : Any other Tikz style option or command can be entered via
  the option style. Most of these commands can be found in the "TikZ and
  PGF Manual". Contain the commands additional options (e.g. shading =
  ball), then the argument for the style has to be between \{ \} brackets.

\item[node\_layer] : the node can be assigned to a specific layer.

\item[node\_label\_off] : is Boolean option which suppress all labels.

\item[node\_label\_as\_id] : is a Boolean option which assigns the node id as label.

\item[node\_math\_mode] : is a Boolean option which transforms the labels into
  mathematical expressions without using the \$ \$ environment.

\item[node\_pseudo] : is a Boolean option which creates a pseudo node, where only
  the node name and the node coordinate will be provided.
\end{description}

\textbf{Keyword arguments for edge styles}

\begin{description}
\item[edge\_width] : width of the edges. The default unit is point (pt).

\item[edge\_color] : color of the edges. The default is gray. Colors can
  be specified either by common color names, or by 3-tuples of floats
  (ranging between 0 and 255 for the R, G and B components).

\item[edge\_opacity] : opacity of the edges. The default is 1. The range of the
  number lies between 0 and 1. Where 0 represents a fully transparent fill
  and 1 a solid fill.

\item[edge\_curved] : whether the edges should be curved. Positive numbers
  correspond to edges curved in a counter-clockwise direction, negative
  numbers correspond to edges curved in a clockwise direction. Zero
  represents straight edges.

\item[edge\_label] : labels drawn next to the edges.

\item[edge\_label\_position] : Per default the label is positioned in between
  both nodes in the center of the line. Classical Tikz commands can be used to
  change the position of the label.

\item[edge\_label\_distance] : The label position between the nodes can be
  modified with the distance option. Per default the label is centered
  between both nodes. The position is expressed as the percentage of the
  length between the nodes, e.g. of distance = 0.7, the label is placed at
  70\% of the edge length away of Vertex i.

\item[edge\_label\_color] : color of the label.

\item[edge\_label\_size] : font size of the label.

\item[edge\_style] : Any other Tikz style option or command can be entered via
  the option style. Most of these commands can be found in the "TikZ and
  PGF Manual". Contain the commands additional options (e.g. shading =
  ball), then the argument for the style has to be between \{ \} brackets.

\item[edge\_arrow\_size] : arrow size of the edges.

\item[edge\_arrow\_width] : width of the arrowhead on the edge.

\item[edge\_loop\_size] :  modifies the length of the edge. The measure value has
  to be insert together with its units. Per default the loop size is 1 cm.

\item[edge\_loop\_position] : The position of the self-loop is defined via the
  rotation angle around the node. The origin (0) is the y axis. A positive
  number change the loop position counter clockwise, while a negative
  number make changes clockwise.

\item[edge\_loop\_shape] : The shape of the self-loop is defined by the enclosing
  angle. The shape can be changed by decreasing or increasing the argument
  value of the loop shape option.

\item[edge\_directed] : is a Boolean option which transform edges to directed
  arrows. If the network is already defined as directed network this option
  is not needed, except to turn off the direction for one or more edges.

\item[edge\_math\_mode] : is a Boolean option which transforms the labels into
  mathematical expressions without using the \$ \$ environment.

\item[edge\_not\_in\_bg] : Per default, the edge is drawn on the background layer
  of the tikz picture. I.e. objects which are created after the edges
  appear also on top of them. To turn this off, the option edge\_not\_in\_bg
  has to be enabled.

\end{description}

\textbf{Keyword arguments for layout styles}

\begin{description}
\item[layout] : dict or string, optional (default = None)
  A dictionary with the node positions on a 2-dimensional plane. The
  key value of the dict represents the node id while the value
  represents a tuple of coordinates (e.g. n = (x,y)). The initial
  layout can be placed anywhere on the 2-dimensional plane.

  Instead of a dictionary, the algorithm used for the layout can be defined
  via a string value. Currently, supported are:

  \begin{description}
  \item[Random layout], where the nodes are uniformly at random placed in the
    unit square. This algorithm can be enabled with the keywords:
    ``Random'', ``random'', ``rand'', or \texttt{None}.

  \item[Fruchterman-Reingold force-directed algorithm]. In this algorithm, the
    nodes are represented by steel rings and the edges are springs between
    them. The attractive force is analogous to the spring force and the
    repulsive force is analogous to the electrical force. The basic idea is
    to minimize the energy of the system by moving the nodes and changing
    the forces between them. This algorithm can be enabled with the
    keywords: ``Fruchterman-Reingold'', ``fruchterman\_reingold'', ``fr'',
    ``spring\_layout'', ``spring layout'', ``FR''.

  \end{description}

  \begin{table}[h]
  \footnotesize%
  \begin{center}
    \begin{tabular}{ll}
      \toprule
      Keys    & Other valid keys  \\
      \midrule
      Random               & Random, random, rand, None                     \\
      Fruchterman-Reingold & Fruchterman-Reingold, fruchterman\_reingold, fr \\
              & spring\_layout, spring layout, FR               \\
      \bottomrule
    \end{tabular}
    \scriptsize
  \end{center}
  \caption{Algorithms keyword naming convention.}
  \label{tab:algo_keywords}
\end{table}

\item[force] : float, optional (default = None)
  Optimal distance between nodes.  If None the distance is set to
  $1/\sqrt{n}$ where $n$ is the number of nodes.  Increase this value to move
  nodes farther apart.

\item[positions] : dict or None  optional (default = None)
  Initial positions for nodes as a dictionary with node as keys and values
  as a coordinate list or tuple.  If None, then use random initial
  positions.

\item[fixed] : list or None, optional (default = None)
  Nodes to keep fixed at initial position.

\item[iterations] : int, optional (default = 50)
  Maximum number of iterations taken

\item[threshold]: float, optional (default = 1e-4)
  Threshold for relative error in node position changes.  The iteration
  stops if the error is below this threshold.

\item[weight] : string or None, optional (default = None)
  The edge attribute that holds the numerical value used for the edge
  weight.  If None, then all edge weights are 1.

\item[dimension] : int, optional (default = 2)
  Dimension of layout. Currently, only plots in 2 dimension are supported.

\item[seed] : int or None, optional (default = None)
  Set the random state for deterministic node layouts. If int, \texttt{seed} is
  the seed used by the random number generator, if None, the a random seed
  by created by the numpy random number generator is used.
\end{description}

\textbf{Keyword arguments for general options}

\begin{description}
\item[units] : string or tuple of strings, optional (default = ('cm','pt'))
  Per default, all size values are based on cm, and all line widths are
  defined in pt units. Whit this option the input units can be
  changed. Currently supported are: Pixel 'px', Points 'pt',
  Millimeters 'mm', and Centimeters 'cm'. If a single value is entered as
  unit all inputs have to be defined using this unit. If a tuple of units
  is given, the sizes are defined with the first entry the line widths with
  the second entry.

\item[layout] : dict
  A dictionary with the node positions on a 2-dimensional plane. The
  key value of the dict represents the node id while the value
  represents a tuple of coordinates (e.g. n = (x,y)). The initial
  layout can be placed anywhere on the 2-dimensional plane.

\item[margins] : None, int, float or dict, optional (default = None)
  The margins define the 'empty' space from the canvas border. If no
  margins are defined, the margin will be calculated based on the maximum
  node size, to avoid clipping of the nodes. If a single int or float is
  defined all margins using this distances. To define different the margin
  sizes for all size a dictionary with in the form of
  \texttt{\{'top':2,'left':1,'bottom':2,'right':.5\}} has to be used.

\item[canvas] : None, tuple of int or floats, optional (default = (6,6))
  Canvas or figure\_size defines the size of the plot. The values entered as
  a tuple of numbers where the first number is width of the figure and the
  second number is the height of the figure. If the option \texttt{units} is not
  used the size is specified in cm. Per default the canvas is $6 \times 6$ cm.

\item[keep\_aspect\_ratio] : bool, optional (default = True)
  Defines whether to keep the aspect ratio of the current layout. If
  \texttt{False}, the layout will be rescaled to fit exactly into the
  available area in the canvas (i.e. removed margins). If \texttt{True}, the
  original aspect ratio of the layout will be kept and it will be
  centered within the canvas.

\item[standalone] : bool, optional (default = True)
  If this option is true, a standalone latex file will be created. i.e. the
  figure can be compiled from this output file. If standalone is false,
  only the tikz environment is stored in the tex file, and can be imported
  in an existing tex file.

\item[clean] : bool, optional (default = True)
  Whether non-pdf files created that are created during compilation should
  be removed.

\item[clean\_tex] : bool, optional (default = True)
  Also remove the generated tex file.

\item[compiler] : str or None, optional (default = None)
  The name of the LaTeX compiler to use. If it is None, cnet will choose a
  fitting one on its own. Starting with \texttt{latexmk} and then \texttt{pdflatex}.

\item[compiler\_args] : list or None, optional (default = None)
  Extra arguments that should be passed to the LaTeX compiler. If this is
  None it defaults to an empty list.

\item[silent] : bool, optional (default = True)
  Whether to hide compiler output or not.

\end{description}

\textbf{Keyword naming convention}

In the style dictionary multiple keywords can be used to address attributes. These keywords will be converted to an unique key word, used in the remaining code. This allows to keep the keywords used in \texttt{igraph}.

\begin{table}[h]\index{Edge!options}
  \footnotesize%
  \begin{center}
    \begin{tabular}{ll}
      \toprule
      Keys    & Other valid keys  \\
      \midrule
      node    & vertex, v, n      \\
      edge    & link, l, e        \\
      margins & margin            \\
      canvas  & bbox, figure\_size \\
      units   & unit              \\
      fixed   & fixed\_nodes, fixed\_vertices, fixed\_n, fixed\_v\\
      positions & initial\_positions, node\_positions, vertex\_positions, n\_positions, v\_positions\\
      \bottomrule
    \end{tabular}
    \scriptsize
  \end{center}
  \caption{Keyword naming convention.}
  \label{tab:keywords}
\end{table}


\backmatter

\bibliography{sample-handout}
\bibliographystyle{plainnat}

\printindex

\end{document}